\documentclass[sigconf,screen]{acmart}
\AtBeginDocument{%
  }
\usepackage{tcolorbox}

\newcommand{\finding}[2]{
\begin{tcolorbox}[width=\linewidth,boxrule=0pt,top=1pt, bottom=1pt, left=1pt,right=1pt, colback=gray!20,colframe=gray!20]
\textbf{Finding #1:} 
{#2}
\end{tcolorbox}}

\copyrightyear{2024}
\acmYear{2024}
\setcopyright{acmlicensed}\acmConference[ASE '24]{39th IEEE/ACM International Conference on Automated Software Engineering }{October 27-November 1, 2024}{Sacramento, CA, USA}
\acmBooktitle{39th IEEE/ACM International Conference on Automated Software Engineering (ASE '24), October 27-November 1, 2024, Sacramento, CA, USA}
\acmDOI{10.1145/3691620.3695004}
\acmISBN{979-8-4007-1248-7/24/10}

\usepackage{subcaption}

\begin{document}

\title{A Systematic Evaluation of Large Code Models in API Suggestion: When, Which, and How}

\author{Chaozheng Wang}
\affiliation{%
  \institution{The Chinese University of Hong Kong}
  \city{Hong Kong}
  \country{China}}
\email{adf111178@gmail.com}

\author{Shuzheng Gao}
\affiliation{%
  \institution{The Chinese University of Hong Kong}
  \city{Hong Kong}
  \country{China}}
\email{szgao23@cse.cuhk.edu.hk}

\author{Cuiyun Gao}
\authornote{Corresponding author. The author is also affiliated with Peng Cheng Laboratory. }
\affiliation{%
  \institution{Harbin Institute of Technology}
  \city{Shenzhen}
  \country{China}}
\email{gaocuiyun@hit.edu.cn}

\author{Wenxuan Wang}
\affiliation{%
  \institution{The Chinese University of Hong Kong}
  \city{Hong Kong}
  \country{China}}
\email{wxwang@cse.cuhk.edu.hk}

\author{Chun Yong Chong}
\affiliation{%
  \institution{Huawei}
  \city{Hong Kong}
  \country{China}}
\email{chunyong@ieee.org}

\author{Shan Gao}
\affiliation{%
  \institution{Huawei}
  \city{Shenzhen}
  \country{China}
  }
\email{gaoshan17@huawei.com}

\author{Michael R. Lyu}
\affiliation{%
  \institution{The Chinese University of Hong Kong}
  \city{Hong Kong}
  \country{China}}
\email{lyu@cse.cuhk.edu.hk}

\begin{abstract}
API suggestion is a critical task in modern software development, assisting programmers by predicting and recommending third-party APIs based on the current context. Recent advancements in large code models (LCMs) have shown promise in the API suggestion task. However, they mainly focus on suggesting which APIs to use, ignoring that programmers may demand more assistance while using APIs in practice including when to use the suggested APIs and how to use the APIs. To mitigate the gap, we conduct a systematic evaluation of LCMs for the API suggestion task in the paper.

To facilitate our investigation, we first build a benchmark that contains a diverse collection of code snippets, covering 176 APIs used in 853 popular Java projects. Three distinct scenarios in the API suggestion task are then considered for evaluation, including (1) ``\textit{when to use}'', which aims at determining the desired position and timing for API usage; (2) ``\textit{which to use}'', which aims at identifying the appropriate API from a given library; and (3) ``\textit{how to use}'', which aims at predicting the arguments for a given API. The consideration of the three scenarios allows for a comprehensive assessment of LCMs' capabilities in suggesting APIs for developers.  During the evaluation, we choose nine popular LCMs with varying model sizes for the three scenarios. We also perform an in-depth analysis of the influence of context selection on the model performance.
Our experimental results reveal multiple key findings. For instance, LCMs present the best performance in the ``how to use'' scenario while performing the worst in the ``when to use'' scenario, e.g., the average performance gap of LCMs between ``when to use'' and ``how to use'' scenarios achieves 34\%, indicating that the ``when to use'' scenario is more challenging.
Furthermore, enriching context information substantially improves the model performance. Specifically, by incorporating the contexts, smaller-sized LCMs can outperform those twenty times larger models without the contexts provided.
Based on these findings, we finally provide insights and implications for researchers and developers, which can lay the groundwork for future advancements in the API suggestion task.
\end{abstract}

\ccsdesc[500]{Software and its engineering}
\ccsdesc[500]{Software and its engineering~Software development techniques}
\keywords{large code models, API suggestion, empirical study}
\maketitle

\section{Introduction}

\begin{figure}
    \centering
    \includegraphics[width=0.49\textwidth]{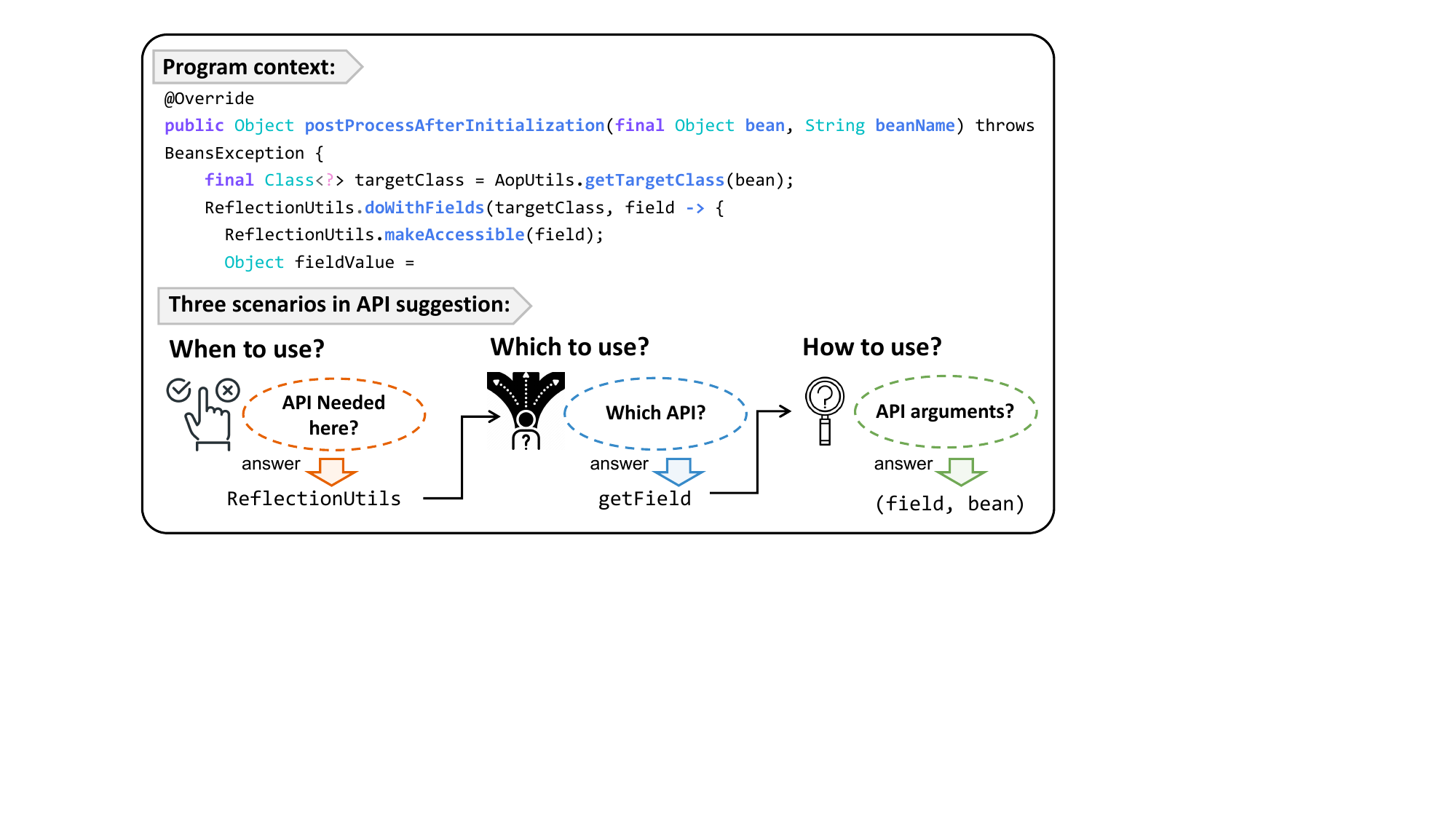}
    \vspace{-12pt}
    \caption{Three distinct scenarios in the API suggestion task.}
    \label{fig:scenario}
    \vspace{-8pt}
\end{figure}


API suggestion is a critical task in modern software development, aiming to assist programmers by predicting and recommending third-party API usage based on the current context \cite{peng2022revisiting, DBLP:journals/tse/ChenGRP0L23, wei2022clear}. With the development of deep learning, multiple techniques have been proposed to provide intelligence API suggestions. In recent years, the emergence of large language models (LLMs) has revolutionized various natural language processing (NLP) tasks \cite{touvron2023llama, Brown2020LanguageMA, xu2023wizardlm}. Inspired by their success, researchers have adapted these models to the domain of programming languages, giving rise to large code models (LCMs) \cite{li2023starcoder, lozhkov2024starcoder, roziere2023code, guo2024deepseek}. LCMs have shown remarkable improvements in the API suggestion task \cite{chen2024apigen, nashid2024contextual}, by leveraging their ability to capture complex patterns and semantics from vast amounts of source code corpus.

Despite the impressive performance of LCMs in the API suggestion task, the previous studies mainly focus on suggesting appropriate APIs from a given library to use \cite{chen2024apigen, DBLP:journals/tse/ChenGRP0L23}, which do not involve the common API usage practices faced by developers as illustrated in Figure \ref{fig:scenario}.
Given the convenience and powerful functionalities of third-party APIs, developers can improve their programming efficiency and productivity. However, the huge amount of APIs requires great effort to memorize. Specifically, developers may demand more assistance while using APIs in practice including when to use APIs and how to use APIs. However, these scenarios of the API suggestion task remain unexplored, leaving a substantial gap in understanding and supporting the diverse needs of developers in real-world programming environments.

\begin{figure*}
    \centering
    \includegraphics[width=0.9\textwidth]{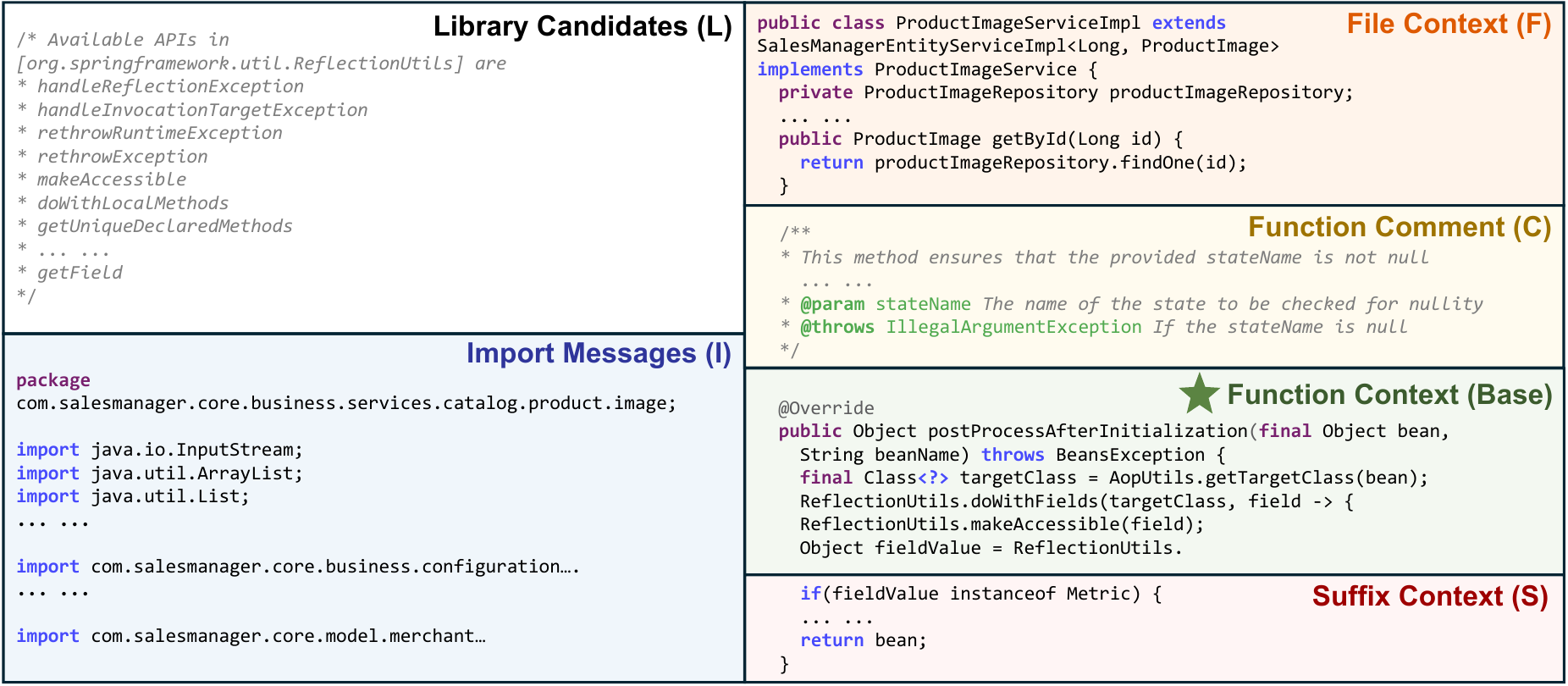}
    \vspace{-6pt}
    \caption{The illustration of different context types.}
    \label{fig:context}
    \vspace{-8pt}
\end{figure*}

In this paper, we present a systematic evaluation of LCMs in API suggestion tasks, aiming at addressing the critical gap in the existing literature. To facilitate the investigation, we first build a benchmark for covering the various API suggestion scenarios. Our benchmark comprises a collection of 4,146 entries of API usage in 3,136 code files, covering 176 diverse APIs sourced from 853 popular Java projects. By curating a wide range of API usage patterns and contexts, we aim to provide a
representative and diverse evaluation dataset for assessing the capabilities of LCMs in real-world API suggestion tasks.

To achieve a comprehensive evaluation, we propose to divide the API suggestion task into three distinct scenarios to simulate the practices of developers: (1) ``\textit{when to use APIs}'', which aims to determine to use APIs in appropriate positions; (2) ``\textit{which API to use}'', which aims to identify the desired API from the given library; and (3) ``\textit{how to use APIs}'', aiming at predicting the arguments for a given API (illustrated in Figure \ref{fig:scenario}). This task decomposition allows us to assess the LCMs' ability to understand and generate code in different API usage scenarios. Furthermore, considering that different contexts provide varying levels of information and guidance, we investigate the impact of various contexts on the performance of LCMs by incorporating different aspects of contexts such as function comments and import messages as shown in Figure \ref{fig:context}. By systematically varying the amount and type of contextual information provided to the LCMs, we aim to identify the most effective prompting strategies for the API suggestion task. To ensure a thorough exploration of the capabilities of LCMs, we include nine LCMs in our experiments: StarCoder~\cite{lozhkov2024starcoder}, CodeLlama~\cite{roziere2023code}, and DeepSeek-Coder~\cite{guo2024deepseek}. These models encompass a wide range of sizes, i.e., from 1 billion to 34 billion parameters, which also enables us to examine the impact of model scales on performance.

Based on our experimental results, we achieve the following key findings: (1)\textbf{ LCMs present the best performance in the ``how to use'' scenario while performing the worst in the ``when to use'' scenario.}
Specifically, the average performance gap of LCMs between ``when to use'' and ``how to use'' scenarios achieves 34\%. ; (2) \textbf{Enriching context information can substantially improve the performance of the API suggestion task.}  For instance, we find that including file contexts, function comments, import messages, and suffix contexts in the input prompt, the average exact match score in the ``when to use'' scenario increases by 89.2\%. 
Specifically, by incorporating additional contexts, smaller-sized LCMs (e.g., DeepSeek-Coder 1.3B) can outperform those twenty times larger models (e.g., DeepSeek-Coder 33B) without the contexts provided.
(3) \textbf{Enriching context information will increase the number of tokens input to LCMs, which consequently decreases the model throughput.} For instance, the average number of tokens will increase more than three times after incorporating all studied
contexts, while the average throughput of LCMs drops by 54\%.
Based on the findings, we finally provide valuable insights into the strengths and limitations of current LCMs in handling API suggestion tasks, which can lay the groundwork for future advancements in the API suggestion task. 


Our main contributions can be summarized as follows:

\begin{itemize}
\item To the best of our knowledge, we are the first to divide the API suggestion task into three scenarios including ``how to use'', ``which to use'', and ``when to use''. The categorization fills a critical gap in the existing literature.

\item We propose an API suggestion benchmark specifically designed to assess the capabilities of LCMs in real-world API usage scenarios.
Based on the benchmark, we conduct extensive experiments with nine popular LCMs for the three scenarios.

\item Based on the results, we finally provide
insights and implications for researchers and developers, which can lay the groundwork for future advancements in the API suggestion task.

\item To foster reproducibility and encourage further research in this area, we make our data and code publicly available at \textit{\url{https://github.com/adf1178/api_suggestion_evaluation}}. 


\end{itemize}

\section{Overview of Methodology}
In this section, we introduce our evaluation methodology from benchmark preparation, context types, and research questions, respectively.

\subsection{Benchmark Preparation}
In this study, we focus exclusively on APIs from the Spring Framework for several key reasons. SpringFramework is one of the most popular frameworks in web development, which is widely used in previous studies \cite{fowkes2016parameter}. Furthermore, it is highly prevalent in Maven repositories \cite{maven}. Its widespread adoption ensures that our benchmark is both relevant and representative of real-world API usage scenarios. In addition, its extensive documentation provides clear and complete descriptions of the APIs, which is essential for comprehensively evaluating the performance of LCMs in API suggestion tasks. By concentrating on a single and well-established framework with thorough documentation, we can conduct a more focused and in-depth analysis.

\subsubsection{Data Collection}
First, we obtain high-star repositories from GitHub by selecting those with more than 100 stars. We then filter out the repositories that do not utilize the Spring Framework, resulting in a final set of 853 projects. Second, for each file in these projects, we use the Tree-sitter~\cite{treesitter} to construct an abstract syntax tree (AST). By traversing the nodes of the AST, we identify function calls and determine if these calls correspond to APIs from the Spring Framework. Third,  based on the frequency of API calls, we prioritize APIs with the highest usage, retaining those that are used more than ten times. This process yields a set of 176 APIs. We then retain the files within the projects that use these high-frequency APIs. To ensure diversity, we select only one file per project for each API, ensuring a varied
dataset. Finally, our benchmark contains 3,136 code files and 4,146 API usage entries.


    

\subsubsection{Benchmark Construction}
After collecting API data, we categorize the API suggestion scenarios into three scenarios including ``\textit{when to use}'', ``\textit{which to use}'', and ``\textit{how to use}'' as shown in Figure \ref{fig:scenario}. These scenarios simulate different situations where developers use APIs in programming practice and evaluate the capabilities of current LCMs to generate accurate suggestions in these scenarios. 

\textbf{When to Use API.}
In the ``when to use'' scenario, we evaluate the capabilities of LCMs to correctly call APIs based on the surrounding code structure and logic. This scenario simulates the situation where developers may be aware of the available APIs but struggle with determining the appropriate timing or location to invoke them within the code. Taking Figure \ref{fig:scenario} as an example, based on the surrounding code, the model should determine that ``\textit{ReflectionUtils.getField(field, bean)}'' is the correct API call to use and place it in the designated location.

\textbf{Which API to Use.}
Which API to use, also called API recommendation in existing studies \cite{DBLP:journals/tse/ChenGRP0L23}, reflects the scenario in which developers input the parent library and the \textit{dot} operator and expect code completion tools to predict the exact API to use. For instance, give ``\textit{ReflectionUtils.}'' and expect LCMs to suggest ``\textit{getField(field, bean)}''. The ``which API to use'' scenario arises when developers have specific functionality in mind but are unsure about the most appropriate API to achieve their goal.

\textbf{How to use APIs.} In this scenario, we simulate the situation where developers input the API and expect LCMs to predict the arguments of the API (API arguments prediction) as illustrated in Figure \ref{fig:scenario}. Due to the increasing number of third-party libraries and their corresponding APIs, developers may struggle to remember how to use the APIs in detail. 



\subsection{Context Types} 
Based on the collected data on API suggestion, we further categorize the contexts in the code file to quantitatively evaluate the influence of different types of contexts on model performance.
Specifically, we utilize the following types of contexts which are demonstrated in Figure \ref{fig:context} to construct the prompt fed into LCMs and evaluate their performance.

\begin{itemize}
    \item \textbf{Function Context (Base)} is the fundamental context used to feed into LCMs for suggesting API usage. Specifically, we select the source code from the function signature of the function involving the target API up to the target API itself. We use function context in our experiments because it represents the basic unit of API usage.

    \item \textbf{File Context (F)} represents the source code outside the function that involves the target API.  In addition to the function context, considering the code contexts at the file level provides LCMs with a broader context, potentially improving their performance in providing API suggestions. It is important to note that a file may contain a substantial amount of source code, which can exceed the context limit of LCMs and result in notable time consumption. Therefore, we select $k$ lines of code preceding the function to construct the file context.

    \item \textbf{Function Comment (C)} typically describes the purpose and logic of the corresponding functions~\cite{garousi2015usage,DBLP:journals/tosem/GaoGHZNXL23}. We incorporate function comments into the context to examine whether explaining the function's utilities in natural language can enhance the performance of LCMs in API suggestion tasks.

    \item \textbf{Suffix Context (S)} refers to the source code following the called API statement. This consideration arises from the fact that developers often edit existing code rather than always appending new code at the end. While current LCM architectures typically generate code tokens auto-regressively (i.e., one token at a time), researchers have proposed fill-in-the-middle (FIM) tasks for pre-training. Through FIM, LCMs can incorporate suffix context to complete intermediate code segments, potentially enriching the model input and enhancing performance in API suggestion tasks.

    \item \textbf{Import Messages (I)} contain the libraries imported in the file and indicate what APIs can be called, which motivates us to experiment with this kind of context as the LCMs' prompt.


    \item \textbf{Library Candidates (L)} provide all of the usable APIs in the currently used parent library, which are designed for the ``which to use'' scenario particularly. We construct this context due to the hallucination issues of LLMs~\cite{rawte2023survey}. In API suggestion scenarios, LCMs may fabricate some APIs that do not exist in the library. Thus, we explore whether explicitly providing usable APIs to LCMs can improve their performance.
\end{itemize}

\subsection{Research Questions}

\subsubsection{RQ1: How do different LCMs perform in the three scenarios of API suggestion?}
In this research question, we divide the API suggestion task into three scenarios, evaluating and comparing the performance of LCMs in each scenario given basic function contexts.

\subsubsection{RQ2: How different types of contexts affect LCMs performance in API suggestion?}
In this research question, we investigate the influence of involving different types of contexts on the model performance in API suggestion. Specifically, we utilize three sub-research questions to investigate the influence of the three scenarios, respectively. 

\subsubsection{RQ3: How do contexts affect the token length and throughput of LCMs?}
More contexts can enrich the semantics of prompts and provide more information but may increase the length of input tokens, which potentially brings overhead to LCMs' inference. Thus, in this RQ, we explore the influence of different types of contexts on the prompt length and throughput of LCMs. 
\begin{table*}[t]
\renewcommand{\arraystretch}{0.85}
    \centering
    \caption{Results of different LCMs in three scenarios of API Suggestion.}
    \vspace{-6pt}
    \begin{tabular}{c|ccccccccc|c}
    \toprule
    Metrics & SC-3B & SC-7B& SC-15B & CL-7B & CL-13B & CL-34B& DSC-1.3B & DSC-6.7B & DSC-33B & Avg \\
    \midrule
    \multicolumn{11}{c}{When to use} \\
    \midrule
    Exact Match & 25.22 &30.49 &31.79 & 30.84& 34.54& 31.52& 23.64& 30.62& \textbf{35.22}& 30.43 \\
    API Acc. & 35.53 & 42.22& 43.22& 42.95& 46.85& 43.99& 34.13& 42.15& \textbf{47.71}& 42.08 \\
    Edit Sim &59.77 & 64.17 & 64.83& 64.63& 66.93& 65.03& 58.89& 63.42& \textbf{67.32}& 63.89 \\
    
    \midrule
    \multicolumn{11}{c}{Which to use} \\
    \midrule
    Exact Match & 50.96 &55.07 & 56.26 & 54.40& 57.20& 53.95& 46.14& 53.79& \textbf{57.22}&53.89 \\
    API Acc. & 77.06 & 80.08& 81.49& 79.47& \textbf{81.83}& 78.77& 71.15& 78.31& 81.29&78.82 \\
    Edit Sim & 81.77& 83.46 &84.11 & 83.25& \textbf{84.33}& 82.80& 79.28& 83.07& 84.12& 82.91 \\
    \midrule
    \multicolumn{11}{c}{How to use} \\
    \midrule
    Exact Match &61.62 &64.66 & 65.34& 64.55& \textbf{66.24}& 64.69& 59.23& 64.21& 65.85&64.04 \\
    Edit Sim & 84.72 &86.27 & 86.50 & 86.15& \textbf{86.88}& 85.91& 83.34& 86.06& 86.69& 85.83 \\
    
    \bottomrule
    \end{tabular}
    
    \label{tab:rq1}
    \vspace{-4pt}
\end{table*}
\section{Experiment Setup}

\subsection{Selected LCMs}
In this paper, we select three kinds of popular and state-of-the-art LCMs
with their versions in different sizes. In specific, our selected LCMs are:
\begin{itemize}
    \item \textbf{StarCoder} \cite{li2023starcoder} is a large language model trained on the mixture of source code and natural language texts. Its training data incorporate more than 80 different programming languages as well as text extracted from GitHub issues and commits and from notebooks.
    We select its 3B, 7B, and 15B versions in our experiments.
    \item \textbf{CodeLlama} \cite{roziere2023code} is a family of large language models for code based on LLama 2~\cite{touvron2023llama} with state-of-the-art code generation, blank infilling, and long-context processing capabilities. In this paper, we choose CodeLlama's base model (i.e., CodeLlama Base) in three different sizes including 7B, 13B, and 34B for instruction tuning. 
    \item \textbf{DeepSeek-Coder} \cite{guo2024deepseek} is a series of large code models that have an identical architecture to CodeLlama. DeepSeek-Coder is trained from 2T tokens from scratch.
    Specifically, we choose DeepSeek-Coder Base in sizes of 1.3B, 6.7B, and 33B in this paper.
\end{itemize}

\subsection{Evaluation Metrics}
In this paper, following previous studies~\cite{svyatkovskiy2020intellicode,tang2023domain,lu2021codexglue,nashid2024contextual}, we utilize three metrics to evaluate the performance of LCMs in the API suggestion tasks including exact match (EM), API usage accuracy, and edit similarity, respectively.
\subsubsection{Exact Match}
The exact match metric measures whether the model output is the same as the ground truth, which is the most strict metric. 

\subsubsection{API Usage Accuracy}
API usage accuracy is utilized to evaluate whether LCMs can predict the desired API in ``which to use'' and ``when to use'' scenarios.

\subsubsection{Edit Similarity} The edit similarity metric is used to measure how closely the model's output resembles the ground truth, considering the edits required to transform one into the other.

\subsection{Implementation Details}
All the experiments are run on a server with 2*A100 GPUs with 80GB graphic memory. For fast inference, we utilize vLLM \cite{kwon2023efficient} based on PagedAttention to improve efficiency. To eliminate the influence of random sampling, we utilize greedy decoding strategy during inference. In addition, we employ the Flash-Attention technique~\cite{dao2022flashattention} for long-context optimization.
\section{Experiment Results}
\begin{table*}[t]
\renewcommand{\arraystretch}{0.9}
    \centering
    \caption{Results in the  ``when to use'' scenario. SC, CL, and DSC indicate StarCoder, CodeLlama, and DeepSeek-Coder, respectively.}
    \vspace{-6pt}
    \begin{tabular}{c|ccccccccc|cc}
    \toprule
    Method & SC-3B & SC-7B& SC-15B & CL-7B & CL-13B & CL-34B& DSC-1.3B & DSC-6.7B & DSC-33B & Avg & Improve\\
    \midrule
    \multicolumn{12}{c}{Exact Match} \\
    \midrule
    Base &25.22 &30.49 &31.79 & 30.84& 34.54& 31.52& 23.64& 30.62& 35.22& 30.43 & \\
    
    Base+F & 27.82 & 32.62 & 34.71& 33.18& 37.65& 34.04& 25.77& 33.39& 38.49&33.07 &($\uparrow8.7\%$)\\
    Base+F+C & 34.90 &40.09 & 41.96& 41.30& 45.41& 41.73& 33.05& 42.51& 46.03& 40.77&($\uparrow34.0\%$)\\
    Base+F+S & 36.11 & 41.75& 43.75& 41.37& 45.89& 45.12& 33.25& 42.83& 48.56&42.07 &($\uparrow38.2\%$)\\
    Base+F+I & 38.36 & 44.21& 48.08& 45.76& 49.12& 47.31& 37.51& 46.07& 50.82& 45.25 &($\uparrow48.7\%$)\\
    Base+F+C+I & 43.45 & 49.39& 52.98& 50.40& 53.40& 51.95& 43.22& 51.15& 55.64&50.17 &($\uparrow64.9\%$)\\
    Base+F+C+I+S & \textbf{52.15} &\textbf{ 58.03}& \textbf{58.68}& \textbf{56.62}& \textbf{59.88}& \textbf{59.33}& \textbf{49.42}& \textbf{59.94}& \textbf{64.08}& \textbf{57.57} &($\uparrow89.2\%$)\\
    \midrule
    \multicolumn{12}{c}{API Usage Accuracy} \\
    \midrule
    Base &35.53 & 42.22& 43.22& 42.95& 46.85& 43.99& 34.13& 42.15& 47.71& 42.08 & \\
    
    Base+F & 38.22 & 44.51& 46.78& 45.63& 50.46&46.56  & 36.28& 45.13& 50.69&44.92&($\uparrow6.9\%$) \\
    Base+F+C & 47.58 & 51.79 & 56.51& 56.09&60.41 & 57.69& 46.07& 56.96& 61.18&54.92 &($\uparrow30.7\%$)\\
    Base+F+S & 46.76 &53.58& 55.35& 53.26& 65.74& 64.99& 43.82& 54.26& 60.68&55.38 &($\uparrow31.8\%$)\\
    Base+F+I & 54.58 &62.57& 64.27& 62.67& 65.74& 64.57& 53.12& 62.27& 67.52&61.92 &($\uparrow47.3\%$)\\
    Base+F+C+I & 60.65 & 66.74 & 69.89& 68.02& 70.83& 69.87& 59.79& 68.27& 72.92&67.44 &($\uparrow60.4\%$)\\
    Base+F+C+I+S & \textbf{67.38} & \textbf{73.03} & \textbf{73.69}& \textbf{72.92}& \textbf{76.39}& \textbf{75.89}& \textbf{65.38}& \textbf{75.35}& \textbf{79.42}& \textbf{73.27}&($\uparrow74.3\%$)\\
    \midrule
    \multicolumn{12}{c}{Edit Similarity} \\
    \midrule
    Base &59.77 & 64.17 & 64.83& 64.63& 66.93& 65.03& 58.89& 63.42& 67.32& 63.89 & \\
    
    Base+F & 61.75 &65.61 & 67.18& 66.26& 69.39&65.44 & 60.48& 65.75& 69.19& 65.67&($\uparrow2.8\%$)\\
    Base+F+C & 67.09 & 70.60& 72.21& 71.71& 74.96& 72.78& 66.24& 72.09& 74.81&71.39 &($\uparrow11.7\%$)\\
    Base+F+S & 68.64 & 72.50& 73.91& 72.54& 75.94& 75.07& 67.11& 73.14& 76.78&72.85 &($\uparrow14.0\%$)\\
    Base+F+I & 70.71 & 74.79& 76.64& 75.94& 77.61& 76.99& 70.08& 75.10& 78.43&75.14&($\uparrow17.6\%$) \\
    Base+F+C+I & 73.95 & 77.56 & 79.41& 78.77& 80.29& 79.52& 73.85 & 78.47& 81.18& 78.11&($\uparrow22.3\%$)\\
    Base+F+C+I+S & \textbf{79.69} & \textbf{83.15} & \textbf{83.58}& \textbf{83.14}& \textbf{85.15}& \textbf{84.11}& \textbf{78.84} & \textbf{84.24}& \textbf{86.49}&\textbf{83.15} &($\uparrow30.1\%$)\\
    \bottomrule
    \end{tabular}
    
    \label{tab:when}
    \vspace{-6pt}
\end{table*}
\subsection{RQ1: Model Performance in API Suggestion}

\begin{table*}[t]
\renewcommand{\arraystretch}{0.9}
    \centering
    \caption{Results in ``which to use'' scenario. SC, CL, and DSC indicate StarCoder, CodeLlama, and DeepSeek-Coder, respectively.}
    \vspace{-6pt}
    \begin{tabular}{c|ccccccccc|cc}
    \toprule
    Method & SC-3B & SC-7B& SC-15B & CL-7B & CL-13B & CL-34B& DSC-1.3B & DSC-6.7B & DSC-33B & Avg & Improve\\
    \midrule
    \multicolumn{12}{c}{Exact Match} \\
    \midrule
    Base & 50.96 &55.07 & 56.26 & 54.40& 57.20& 53.95& 46.14& 53.79& 57.22&53.89 & \\
    
    Base+F & 53.18 & 56.67 &58.34 & 57.09& 59.40& 56.42& 48.43& 56.08& 59.49&56.12 &($\uparrow4.1\%$)\\
    Base+F+L & 53.09 & 57.33 & 58.86& 57.65& 59.68& 56.59& 49.61& 57.44& 59.74 & 56.67 &($\uparrow5.2\%$)\\
    Base+F+C & 58.00 &61.46 & 62.33& 61.52& 63.59& 61.05& 54.45& 61.37& 64.14& 60.88&($\uparrow13.0\%$)\\
    Base+F+S & 60.26 & 63.32 & 65.48& 61.49& 64.42& 62.50& 50.69& 60.49& 65.86 & 61.61 &($\uparrow14.3\%$)\\
    Base+F+I & 55.25 & 60.05& 62.12& 59.63& 62.96& 61.22& 53.41& 60.12& 63.08& 59.76&($\uparrow10.9\%$)\\
    Base+F+C+I & 58.84 & 63.69& 65.53 & 63.91& 66.36& 64.86& 58.05& 64.61& 67.53&63.71 &($\uparrow18.2\%$)\\
    Base+F+C+I+S & \textbf{66.40} & \textbf{69.92}& \textbf{71.77}& \textbf{67.53}& \textbf{70.26}& \textbf{68.69}& \textbf{59.11}& \textbf{68.19}& \textbf{72.69}& \textbf{68.28}&($\uparrow26.7\%$)\\
    Base+F+L+C+I+S & 64.85 & 68.46& 71.22& 66.64& 69.55& 68.21& 58.24& 67.51& 71.94& 67.40&($\uparrow25.1\%$)\\
    \midrule
    \multicolumn{12}{c}{API Usage Accuracy} \\
    \midrule
    Base & 77.06 & 80.08& 81.49& 79.47& 81.83& 78.77& 71.15& 78.31& 81.29&78.82& \\
    
    Base+F & 79.00 & 81.35& 82.62& 80.84& 83.31&80.91  & 73.18& 80.07& 82.91&80.46 &($\uparrow2.1\%$) \\
    Base+F+L & 79.34 & 82.26 &83.85 & 81.89&83.66 & 82.07& 75.22& 82.56& 83.72& 81.62&($\uparrow3.5\%$)\\
    Base+F+C & 84.49 & 86.43 &87.18 & 86.03&87.92 & 85.86& 80.32& 86.15& 87.96&85.81 &($\uparrow8.9\%$)\\
    Base+F+S & 81.17 &82.53& 85.11& 82.16& 86.46& 84.75& 71.56& 80.68& 85.79& 82.24 &($\uparrow4.3\%$)\\
    Base+F+I & 82.12 &84.90& 85.82& 83.64& 86.46& 84.77& 79.03& 84.14& 85.90&84.08 &($\uparrow6.7\%$)\\
    Base+F+C+I & 85.54 & 88.31 & 89.39& 87.58& 89.56& 88.20& 83.34& 87.94& 90.05& 87.77&($\uparrow11.3\%$)\\
    Base+F+C+I+S & \textbf{87.20} & \textbf{88.90} & \textbf{90.52}& \textbf{87.66}& \textbf{89.73}& \textbf{89.11}& \textbf{81.68}& \textbf{88.07}& \textbf{91.29}& \textbf{88.24}& ($\uparrow12.0\%$)\\
    Base+F+L+C+I+S & 84.68 & 86.71 & 89.24& 86.54& 88.75& 88.69& 80.82& 87.75& 90.86& 87.12&($\uparrow10.5\%$)\\
    \midrule
    \multicolumn{12}{c}{Edit Similarity} \\
    \midrule
    Base & 81.77& 83.46 &84.11 & 83.25& 84.33& 82.80& 79.28& 83.07& 84.12& 82.91 & \\
    
    Base+F & 83.07 &84.45 & 85.15& 84.59& 85.38&84.28 & 80.51& 84.18& 85.40& 84.12&($\uparrow1.5\%$)\\
    Base+F+L & 82.99 & 84.72& 85.42& 84.96& 85.65& 84.64& 81.21& 84.89& 85.72&84.47&($\uparrow1.9\%$) \\
    Base+F+C & 85.33 & 86.45& 87.09& 86.67& 87.47& 86.48& 83.59& 86.73& 87.66&86.39&($\uparrow4.2\%$) \\
    Base+F+S & 85.74 & 86.54& 87.68& 86.44& 87.40& 86.61 & 81.53& 85.90& 88.08&86.21&($\uparrow4.0\%$) \\
    Base+F+I & 84.60 & 86.13& 86.90& 86.11& 87.16& 86.52 & 83.35& 86.14& 87.31&86.02 &($\uparrow3.7\%$)\\
    Base+F+C+I & 86.14 & 87.68 & 88.52& 87.88& 88.75& 88.16& 85.28 & 87.96& 89.22&87.73 &($\uparrow5.8\%$)\\
    Base+F+C+I+S & \textbf{88.62} & \textbf{89.70} & \textbf{90.42}& \textbf{89.23}& \textbf{90.04}& \textbf{89.74}& \textbf{85.58}& \textbf{89.44}& \textbf{91.05}&\textbf{89.31} & ($\uparrow7.7\%$)\\
    Base+F+L+C+I+S & 87.58 & 89.19 & 90.20& 88.82& 89.87& 89.05& 85.17& 89.29& 90.88&88.89 &($\uparrow7.2\%$)\\
    \bottomrule
    \end{tabular}
    
    \label{tab:which}
\end{table*}
We present the results of nine LCMs, given the basic function context, in the three scenarios of API suggestion in Table \ref{tab:rq1}. From the table, we achieve the following observations.

\textbf{(1) The model performance increases as the complexity of the scenarios decreases}. For the three scenarios studied in this paper, the ``when to use'' scenario is the most challenging, and ``how to use'' is the simplest one.
We estimate the difficulties of different scenarios based on the fact that ``how to use'' only requires models to predict the API arguments, while the ``which to use'' and ``when to use'' scenarios require further prediction of the specific API (e.g., \textit{``getField''}) and its library (e.g., \textit{``ReflectionUtils''}), respectively. 

From
table~\ref{tab:rq1}, we find that in the most challenging ``when to use'' scenario, which requires LCMs to determine whether to use APIs by themselves, the models achieve an average exact match rate of 30.43. For the relatively simpler ``which to use'' and ``how to use'' scenarios, the average exact match scores for nine LCMs increase to 53.89 and 64.04, representing the improvement of 77\% and 110\%, respectively.


\textbf{(2) Model performance is positively correlated to the model sizes, and the correlation is more pronounced in more challenging scenarios}. 
During evaluating the performance of various LCMs, we observe a clear positive correlation between model sizes and performance. For instance, in the ``when to use'' scenario, the Pearson correlation coefficient between model size and the exact match score achieves 0.67 with a p-value of 0.049, indicating a significant relationship. This positive correlation can be attributed to larger models' enhanced capacity to understand context, resulting in more accurate completions. However, it is important to note that larger models do not consistently outperform their smaller counterparts. Among the tested LCMs with basic context, DeepSeek-Coder with 33B parameters achieves the highest performance in the ``when to use'' scenario, and CodeLlama 13B performs the best in the ``which to use'' and ``how to use'' scenarios. 

As the complexity of scenarios decreases, the relationship becomes less pronounced accordingly. Specifically, the correlation coefficients between model sizes and exact match score decrease to 0.53 and 0.32 in the ``which to use'' and ``how to use'' scenarios, respectively. Such a relationship is also reflected in the performance gap between large and small models. For instance, in the ``when to use'' scenario, DeepSeek-Coder 33B outperforms the 1.3B version by 49\% in terms of the exact match score. However, the difference is narrowed to 24\% and 11\% in the ``which to use'' and ``how to use'' scenarios, respectively. Consequently, we conclude that the impact of model size on performance becomes more substantial as the complexity of the scenarios increases.


\vspace{-6pt}
\finding{1}{The model performance increases as the complexity of the scenarios decreases, i.e., the gap of average exact match score between ``when to use'' and ``how to use'' scenarios achieves 34\%. In addition, model performance is positively correlated to the model sizes, and the correlation is more pronounced in more challenging scenarios.}
\vspace{-6pt}

\subsection{RQ2: Influence of Contexts on Effectiveness}\label{sec:rq2}
In this research question, we investigate the influence of contexts on model performance in each scenario, respectively.

\subsubsection{RQ2.1 When to Use}
We
evaluate LCMs in the ``when to use'' scenario, with the results shown in Table \ref{tab:when}.

From the table, we can find that LCMs achieve an average of 30.43, 42.08, and 63.89 in exact match, API Usage Accuracy, and edit similarity metrics given basic function
context (Base), respectively. These results indicate that current LCMs struggle to effectively suggest the use of APIs in the desired positions with only local function contexts. Besides function context, pretending file context (F) contributes to LCMs' performance (Base+F), improving the three metrics by 8.7\%, 6.9\%, and 2.8\%, respectively. Such results suggest that the file context provides additional information that helps LCMs more accurately complete API calls. Note that due to the potential sizes of files, we include ten lines of code before the function as the file context \cite{liu2024non}. We further explore the influence of varying amounts of file
context on model performance in Section \ref{sec:select}.

Based on file
contexts, we further explore the influence of code comments (+C), import messages (+I), and suffix contexts (+S), respectively. From the table, we observe that these three contexts can further improve the performance of LCMs in API suggestion, i.e., the exact match is increased by 34.0\%, 48.7\%, and, 38.2\% compared to that with basic function context, respectively. We attribute the improvement of adding code comments to that comments reflect the code's functionalities and thus provide more guidance for using the APIs. 
With import messages that define the parent libraries to use in the file, LCMs can better understand the intent of current files and fill in API arguments that conform to the development requirements. In addition, the broad generation space in the ``when to use'' scenario requires LCMs to determine whether to use APIs. Therefore, import messages improve the performance by a large margin, which is the largest improvement among the three kinds of contexts.
As observed from the table, the suffix contexts also contribute substantially to the performance, i.e., the exact match is increased by 38.2\%. Such improvement can be attributed to that 1): involving the suffix contexts ensures contextual coherence: The suffix context provides the expected direction of the code, allowing the model to better understand the current code's logic and structure, leading to more reasonable completions; and 2) suffix context helps to resolve ambiguities present in the earlier parts of the code, providing clear context and making the model's predictions more precise and consistent. We provide a case study to further demonstrate the influence of suffix contexts in Section \ref{sec:case}.

Furthermore, we investigate the impact of combining different types of contexts (i.e., rows Base+F+C+I and Base+F+C+I+S). We can observe that involving both comments and import messages further increases the performance of LCMs, i.e., the three evaluation metrics are increased by 64.9\%, 60.4\%, and 30.1\%, respectively. After combining comments, import messages, and suffix contexts with file contexts, LCMs obtain the most performance gain by nearly 90\% (the exact match rate is improved from 30.43 to 57.57). In addition, we observe that incorporating all studied contexts, small-sized LCMs can outperform those twenty times larger models with the basic function context. For instance, DeepSeek-Coder 1.3B with Base+F+C+I+S contexts outperforms DeepSeek-Coder 33B  with basic function context by 40\% in terms of the exact match score.

The results demonstrate that different types of contexts contribute variably to API suggestion, and enriching prompts by combining these contexts further enhances the performance of LCMs in suggesting APIs.

\begin{table*}[t]
\renewcommand{\arraystretch}{0.9}
    \centering
    \caption{Results in the ``how to use'' scenario. SC, CL, and DSC indicate StarCoder, CodeLlama, and DeepSeek-Coder, respectively.}
    \vspace{-6pt}
    \begin{tabular}{c|ccccccccc|cc}
    \toprule
    Method & SC-3B & SC-7B& SC-15B & CL-7B & CL-13B & CL-34B& DSC-1.3B & DSC-6.7B & DSC-33B & Avg & Improve\\
    \midrule
    \multicolumn{12}{c}{Exact Match} \\
    \midrule
    Base &61.62 &64.66 & 65.34& 64.55& 66.24& 64.69& 59.23& 64.21& 65.85&64.04 &  \\
    
    Base+F & 63.80 & 66.24& 67.20& 66.70& 68.06& 66.43& 60.83& 66.20& 68.08&65.95 &($\uparrow3.0\%$) \\
    Base+F+C & 65.81 & 68.43 & 69.23& 68.94& 70.24 & 68.89& 63.93& 68.74& 70.64&68.31 &($\uparrow6.7\%$) \\
    Base+F+S & 69.35 & 72.07& 73.55& 70.52& 72.11& 71.38& 64.32& 70.21& 73.24& 70.74&($\uparrow10.5\%$)\\
    Base+F+I & 63.65 & 68.04& 69.48& 68.28& 70.15& 69.37& 63.12& 68.52& 70.48& 67.90&($\uparrow6.1\%$)\\
    Base+F+C+I & 65.69 & 70.04& 71.30& 70.68& 72.20& 71.33& 65.67& 71.02 & 73.03& 70.10&($\uparrow9.5\%$)\\
    Base+F+C+I+S & \textbf{72.24} & \textbf{75.37}& \textbf{76.26}& \textbf{74.13}& \textbf{75.37}& \textbf{74.89}& \textbf{67.85}& \textbf{74.11} & \textbf{77.54}& \textbf{74.19}&($\uparrow15.8\%$)\\
    \midrule
    \multicolumn{12}{c}{Edit Similarity} \\
    \midrule
    Base &84.72 &86.27 & 86.50 & 86.15& 86.88& 85.91& 83.34& 86.06& 86.69& 85.83&\\
    
    Base+F & 85.82 & 87.21& 87.59& 87.24& 87.79&86.96 & 84.35& 87.21& 87.83&86.89&($\uparrow1.6\%$) \\
    Base+F+C & 86.83 & 88.07& 88.45& 88.17& 88.76& 88.22& 85.90& 88.09& 88.98& 87.94&($\uparrow2.8\%$)\\
    Base+F+S & 88.19 & 89.60 & 90.18& 88.59& 89.66& 89.12& 85.79& 88.70& 90.15&88.82&($\uparrow3.8\%$) \\
    Base+F+I & 86.40 & 87.83 & 88.68& 88.16& 88.84& 88.52& 85.75& 88.25& 89.03&87.94&($\uparrow2.8\%$) \\
    Base+F+C+I & 87.25 & 88.81& 89.34& 89.14& 89.69 & 89.46& 86.72 & 89.28& 90.26&88.88&($\uparrow3.9\%$) \\
    Base+F+C+I+S & \textbf{89.39} & \textbf{90.76}& \textbf{91.29}& \textbf{90.17}& \textbf{91.11} & \textbf{90.50}& \textbf{87.33} & \textbf{90.17}& \textbf{91.81}&\textbf{90.16}&($\uparrow5.4\%$) \\
    \bottomrule
    \end{tabular}
    
    \label{tab:how}
\end{table*}

\subsubsection{RQ2.2: Which to Use}

The results of LCMs in the ``which to use'' scenario of API suggestion are shown in Table \ref{tab:which}.

Similarly, the improvement obtained by enriching contexts is also observed in this scenario. For instance, including file contexts (+F) and equipped with additional information such as comments and suffix contexts (+C and +S) present 13\% and 14\% improvement in terms of the exact match metric, respectively. In addition, we also experiment with a unique type of context in the ``which to use'' scenario, i.e., library candidates (+L). From the table, we find that incorporating library candidates (+F+L) further improves the performance of file contexts (+F) by 1.1\%, 1.4\%, and 0.4\% on the exact match, API usage accuracy, and edit similarity, respectively. The improvement suggests that informing LCMs of available APIs is helpful in predicting APIs more accurately. However, when involving all of the contexts (+F+L+C+I+S), LCMs perform worse than the combination without library candidates (+F+C+I+S), i.e., 67.40 and 68.28 in the exact match score. The difference indicates that when involving enough contexts, information about library candidates may become redundant.

\subsubsection{RQ2.3: How to Use}

We investigate how LCMs perform in the ``how to use'' scenario (i.e., the ability to fill an API's arguments), and the results are presented in Table \ref{tab:how}.

Regarding the influence of different contexts, our results indicate that the contribution of contexts in the ``how to use'' scenario is less substantial compared to that in the other scenarios. For instance, after involving all available contexts (+F+C+I+S), the exact match and edit similarity are improved by 15.8\% and 5.4\% compared with function context, respectively. These results suggest that, despite the naturally higher accuracy in the ``how to use'' scenario, LCMs derive less additional benefit from the contexts. We hypothesize that in simpler scenarios, LCMs already achieve high accuracy with minimal context, thus additional context provides relatively less guidance and improvement compared to that in more complex scenarios where the LCMs benefit more from enriched contextual information.


\finding{2}{Enriching contexts can substantially enhance the performance of the API suggestion task, with improvements becoming more pronounced as the complexity of scenarios increases. Furthermore, equipped with all studied contexts, smaller-sized LCMs can outperform more than twenty times larger models when no additional context is provided.}
\vspace{-6pt}

\begin{figure}
    \centering
    \begin{subfigure}[b]{0.49\textwidth}
        \centering
        \includegraphics[width=1\textwidth]{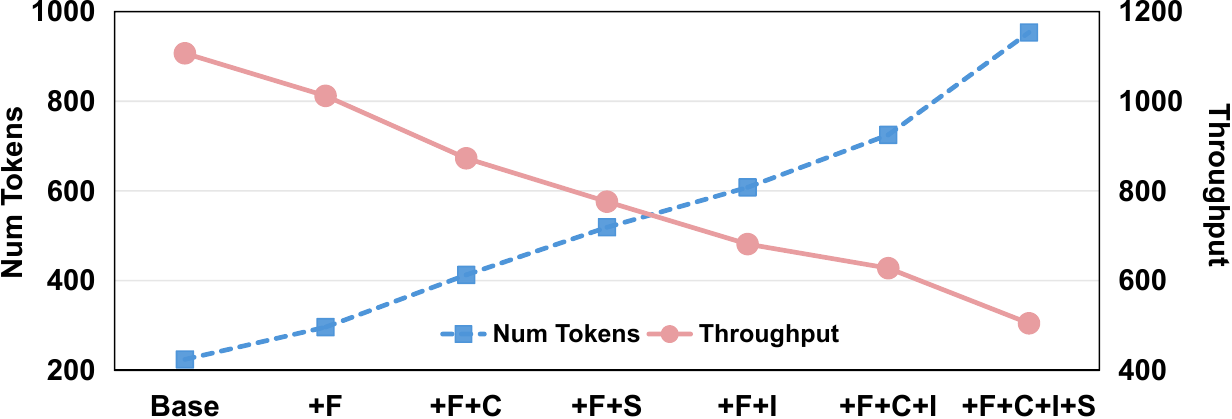}
        \caption{Token numbers and average throughput of different contexts.}
    \end{subfigure}
    \hfill
    \begin{subfigure}[b]{0.49\textwidth}
        \centering
        \includegraphics[width=1\textwidth]{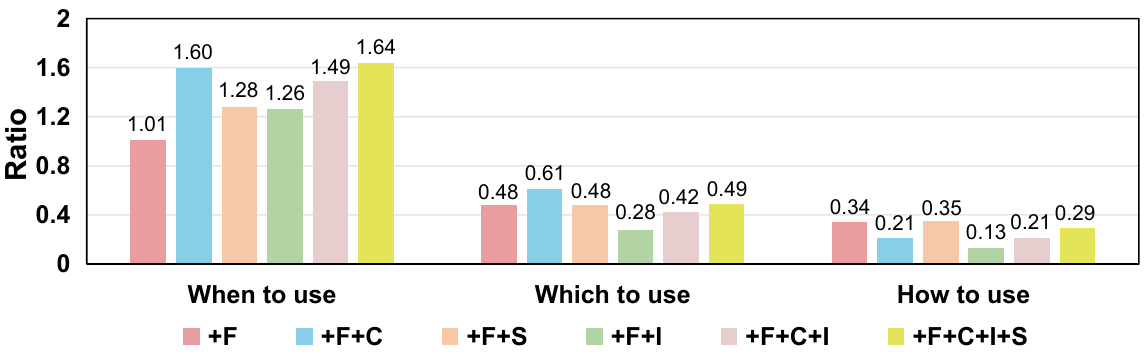}
        \caption{Avg. Exact Match Improvement / Avg. Throughput Decrease ratio of different contexts  in different scenarios.}
    \end{subfigure}
    \caption{Average token length, throughput, and the ratio between performance and throughput.}
    \label{fig:efficiency}
    \vspace{-0.4cm}
\end{figure}

\begin{figure*}
  \begin{subfigure}{0.33\textwidth}
    \includegraphics[width=\textwidth]{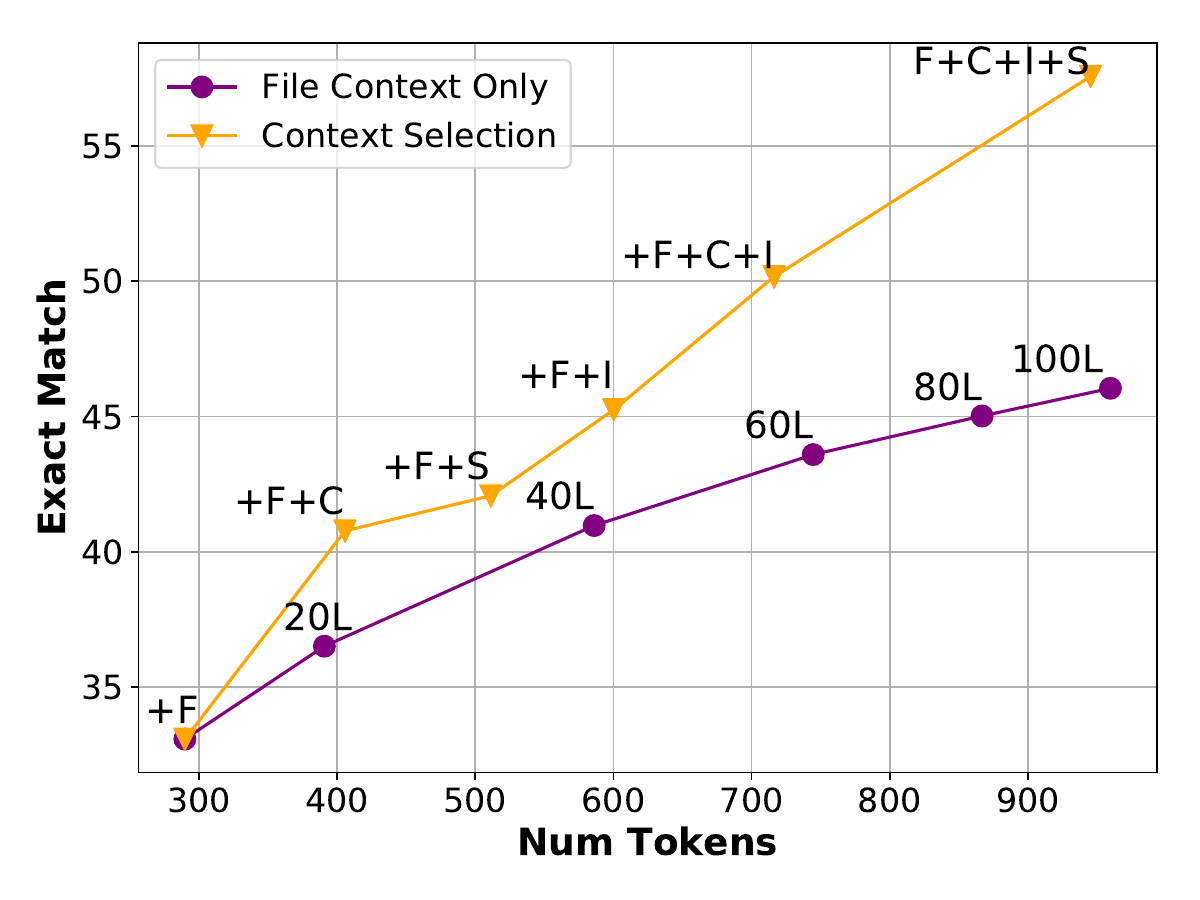}
    \caption{When to use}
  \end{subfigure}%
  \begin{subfigure}{0.33\textwidth}
    \includegraphics[width=\textwidth]{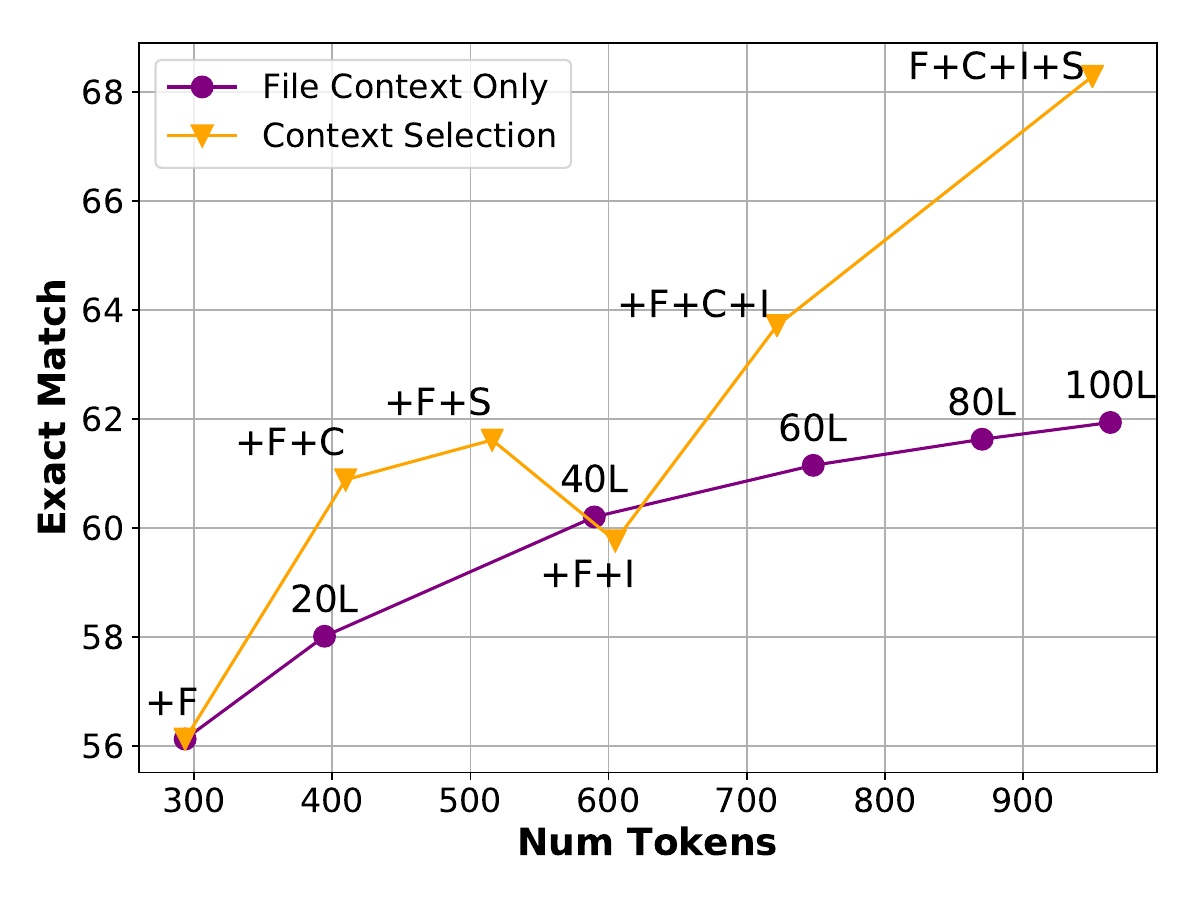}
    \caption{Which to use}
  \end{subfigure}%
  \begin{subfigure}{0.33\textwidth}
\includegraphics[width=\textwidth]{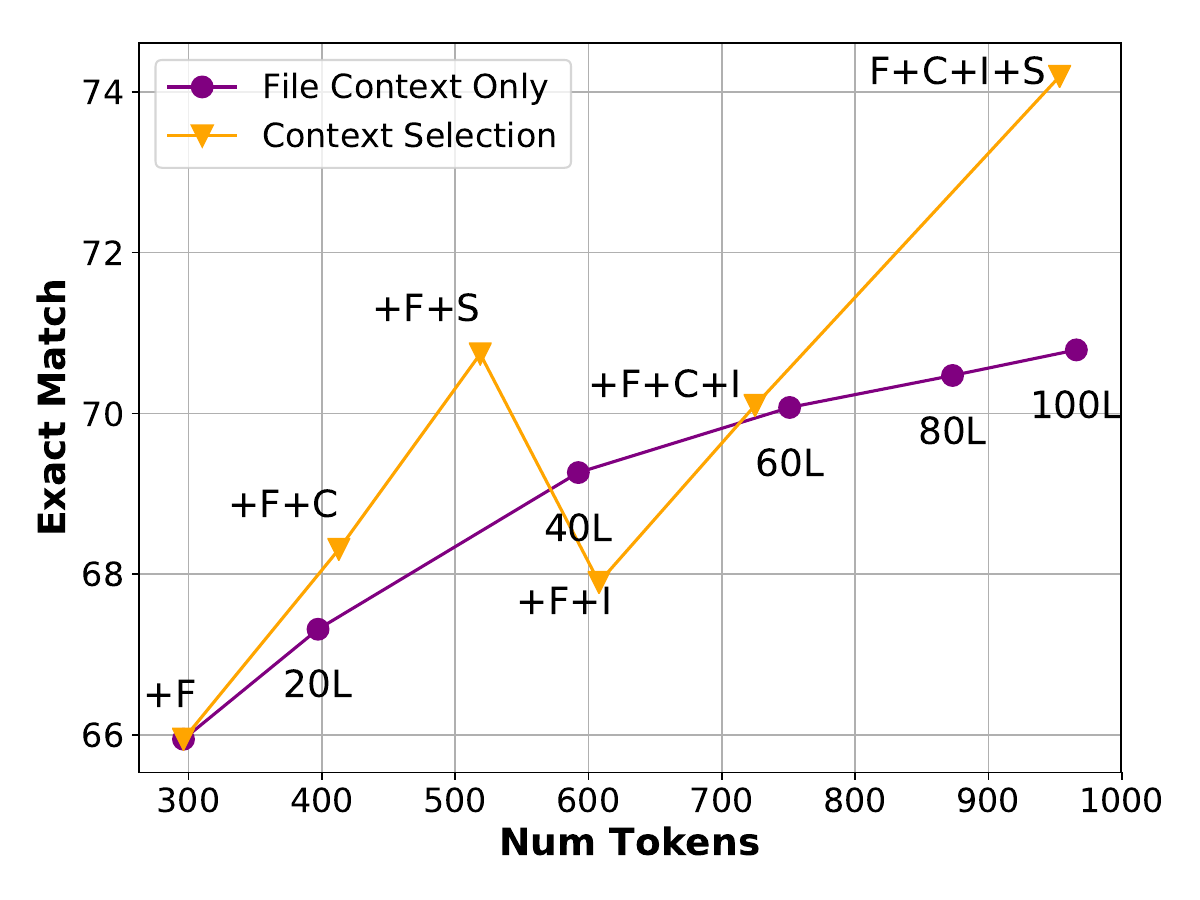}
    \caption{How to use}
  \end{subfigure}%
  \vspace{-8pt}
  \caption{Performance and token numbers 
  of different contexts. ``L''denotes the number of lines included in the file context.}
  \label{fig:file}
\end{figure*}

\subsection{RQ3: Influence of Contexts on
Efficiency}
Based on the results in RQ2, we observe that incorporating contexts enhances the model performance. However, they also increase the prompt length, possibly leading to higher latency. To comprehensively investigate the influence of involving different contexts on models' performance, we explore how different contexts influence the lengths of input tokens.  In addition, we also explore the model efficiency by recording the model throughput under different context settings. The throughput is defined as the average number of tokens that LCMs generate per second \cite{kwon2023efficient, papaioannou2024importance, agrawal2024taming}. The results are presented in Figure \ref{fig:efficiency}. From the figure, we achieve the following observations.

\textbf{(1) Enriching contexts increases the prompt length, which consequently decreases the model throughput}. Figure \ref{fig:efficiency} (a) presents the average token lengths
of input prompts with different contexts. As described in the figure, the token length increases progressively with the addition of each context type. The basic context has the shortest token length (217.8 tokens), while the combination of all context types (+F+C+I+S) results in the longest token length, nearly 1,000 tokens. Consequently, the average throughput decreases by up to 54.4\%. These results suggest that while combining different contexts can enhance API suggestions, it also incurs a trade-off by negatively impacting throughput to some extent.

\textbf{(2) The average improvement gain per unit of throughput decrease differs along with
the context types used in each scenario.}
Enriching contexts introduces a consistent throughput overhead across the three scenarios we experiment with; however, the impacts vary as shown in Figure \ref{fig:efficiency} (b). 
For instance, combining file contexts and suffix contexts (+F+S) yields the best trade-off in the ``how to use'' scenario, where the exact match score improves the most per unit decrease in throughput.
In the ``which to use'' scenario, including file contexts and code comments (+F+C) achieves the best improvement-to-throughput ratio. For the most complex ``when to use'' scenario, we observe that involving all types of contexts (+F+C+I+S) provides the best trade-off. 
Moreover, we observe that in both the ``how to use'' and ``which to use'' scenarios, file contexts and import messages (+F+I) contribute the least, with ratios of only 0.13 and 0.28, respectively. This poor trade-off results from the substantial amount of tokens added by including import messages, while the improvement in these two scenarios remains limited. However, in the ``when to use'' scenario, this combination improves the suggestion accuracy by 48.71\%, making the improvement-to-throughput ratio 25\% higher than file contexts (+F) at 1.26.

The differences among the scenarios indicate that each scenario requires a distinct context selection approach to achieve the optimal trade-off between model performance and throughput.
\vspace{-4pt}
\finding{3}{Additional contexts increase the amount of tokens in the input. Compared to basic function context, incorporating all studied contexts increases the token length by 335\%, which consequently decreases the average throughput by 54\%. In addition, for different scenarios, the context selection approach that achieves the best trade-off between model performance and throughput is also different.}
\vspace{-10pt}
\section{Discussion}

\subsection{Analysis of Context Selection}\label{sec:select}
In the results presented in Section \ref{sec:rq2}, we construct file contexts by including ten lines of source code preceding the function of the target API. This setting is adopted for dealing with larger-sized single files,
which could exceed the token capacity of current LCMs (e.g., DeepSeek-Coder's limit of 8,192 tokens). Previous research \cite{he2021pyart, ciniselli2021empirical} has shown that additional context before the suggestion position can provide more information for model prediction. In this section, we conduct experiments to compare the effects of simply involving more file contexts with our context selection approaches, i.e., selecting contexts with various types.

Specifically, besides the setting used in Section \ref{sec:rq2} that file context (+F) involves ten lines of code preceding the local function (i.e., basic function context), we further evaluate LCMs with extended file contexts of 20, 40, 60, 80, and 100 lines. These settings are chosen due to their comparable context token length to our context selection approaches i.e., the token length varies from around 300 to 950, ensuring that the influence of token length on model performance is eliminated in the comparison. The comparison of these settings is presented in Figure \ref{fig:file}.

As depicted in the figure, extending file contexts enhances API suggestion performance across all three scenarios. Specifically, when using 100 lines of code to serve as file contexts (averaging about 960 tokens in the input prompt), the exact match scores improve by 7.3\%, 10.4\%, and 39.2\% in the respective scenarios compared to using only ten lines of file context. This remarkable improvement demonstrates that LCMs benefit from extended file
contexts in API suggestion tasks. However, we can observe that as the number of tokens increases, the performance improvement becomes slower. This indicates that the benefit of simply introducing longer text without selection is limited. 

Despite the obvious improvements achieved by extending file contexts, our context selection approach yields better suggestion accuracy while using tokens with similar lengths.
As illustrated in the figure, the context selection curve is generally positioned in the upper left of the file-level contexts curve. Despite a context selection (+F+I) performing worse than file contexts with a similar amount  (about 600) of tokens in ``how to use'' and ``which to use'' scenarios, we still have other selection choices to achieve better performance with fewer tokens in the input (e.g., +F+S). 
These results indicate that a carefully curated selection of contexts can outperform the simple extension of file-level contexts with the same or even smaller token length.

Therefore, we conclude that increasing the range of file contexts also benefits the API suggestion task. However, selecting contexts from various types proves to be a more efficient and effective strategy for the API suggestion task, i.e., achieving better performance with the same or even fewer input tokens.

\subsection{Case Study}\label{sec:case}
\begin{figure}
    \centering
    \includegraphics[width=0.49\textwidth]{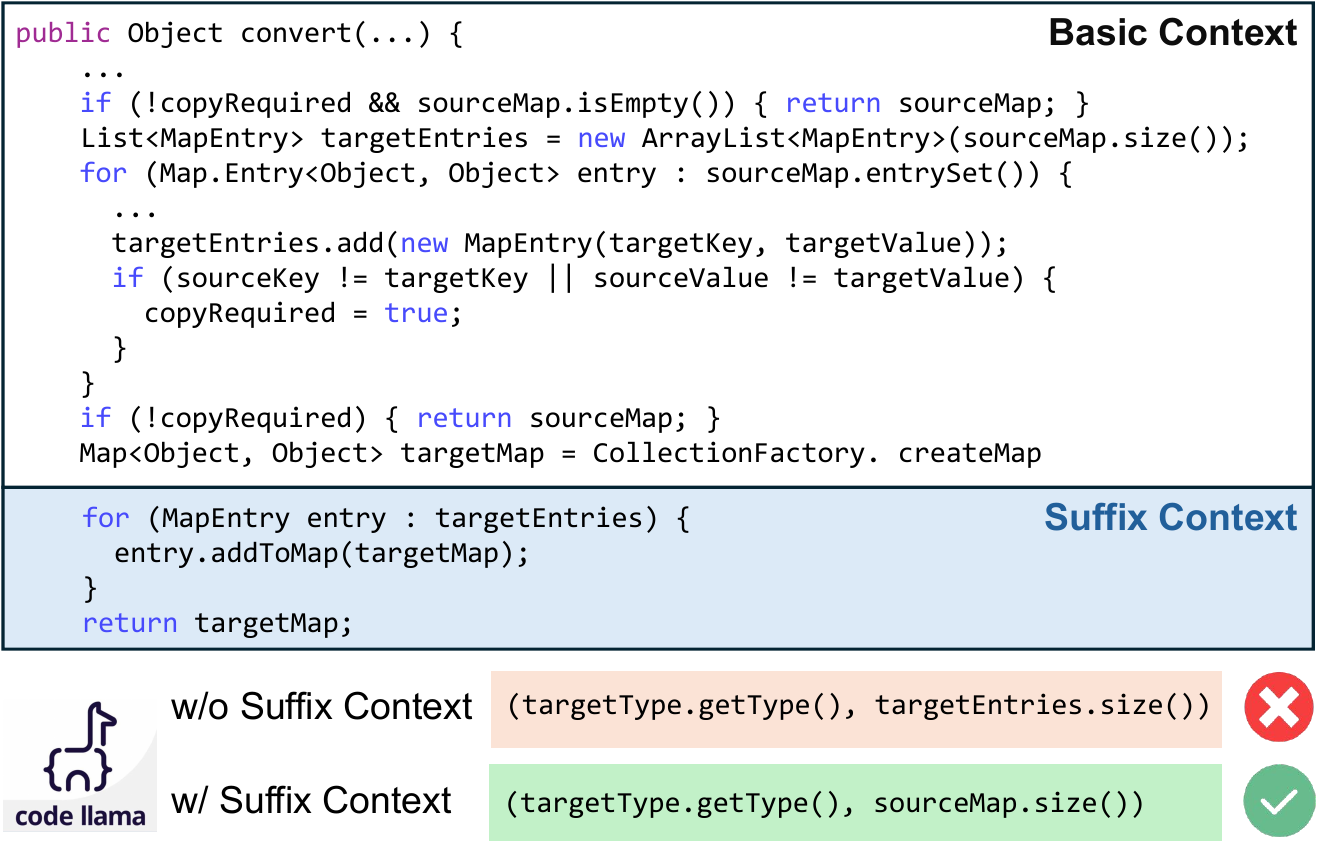}
    \vspace{-8pt}
    \caption{Case study in the ``how to use'' scenario, where the experimented LCM is CodeLlama 7B.}
    \label{fig:case_suffix}
    \vspace{-10pt}
\end{figure}
\begin{figure}
    \centering
    \includegraphics[width=0.485\textwidth]{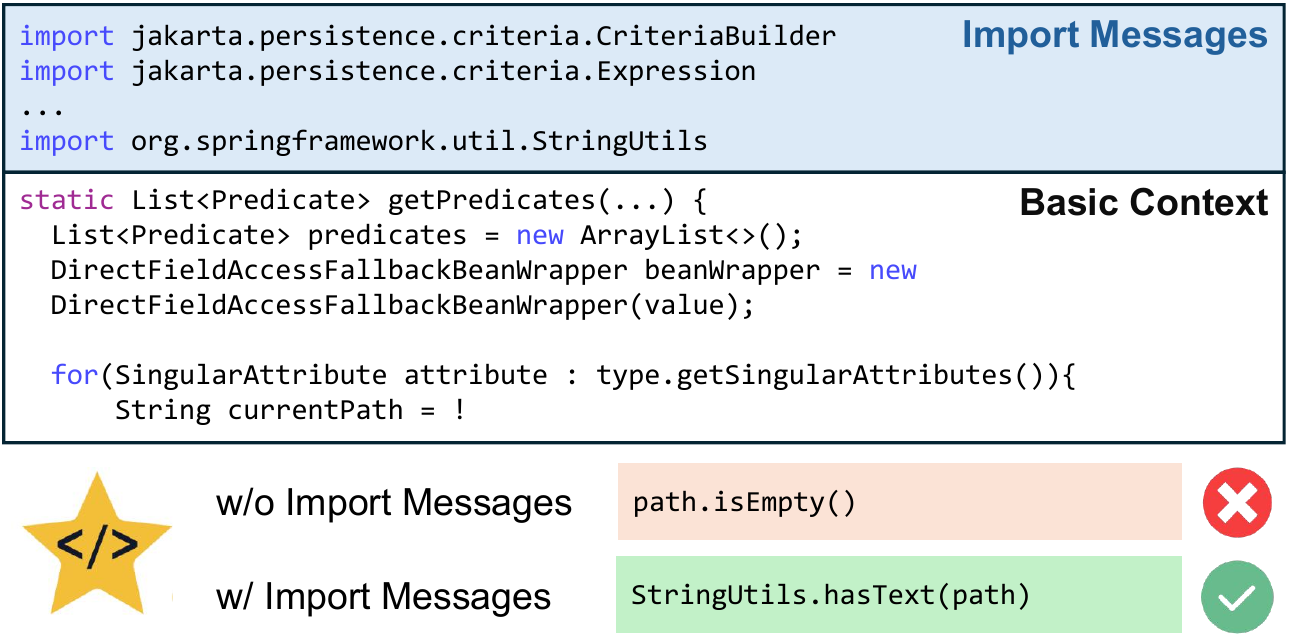}
    \vspace{-8pt}
    \caption{Case study in the ``when to use'' scenario, where the experimented LCM is StarCoder 15B.}
    \label{fig:case_import}
    \vspace{-12pt}
\end{figure}
In this section, we conduct a case study to further illustrate the improvement brought by various contexts. The example case is presented in Figure \ref{fig:case_suffix}. From the figure, we observe that CodeLlama 7B predicts the incorrect second argument as ``\textit{targetEntries.size()}'' when only provided with the basic function context. We hypothesize that because the newly defined object is named ``\textit{targetMap}'', the model mistakenly associates its size with ``\textit{targetEntries}''. However, after incorporating the suffix context, the model successfully suggests the correct usage of the API. We attribute this improvement to the loop in the suffix context, which indicates that all entries from ``\textit{sourceMap}'' eventually populate ``\textit{targetMap}''. This helps the model recognize that ``\textit{targetEntries}'' is not the source of the initial map size. 

We also provide a case study in the ``when to use'' scenario, where the example is presented in Figure \ref{fig:case_import}. From the figure, we find that with basic context, StarCoder 15B defaults to a common method available on the ``\textit{path}'' object and calls ``\textit{isEmpty()}'', a widely known method to check if a string is empty. Contrarily, involving import messages facilitates the model in successfully predicting the desired API calls. We suppose that, after incorporating import messages, the model recognizes that ``\textit{StringUtils}'' provides utility methods for string operations, and then uses the ``\textit{hasText}'' API.

\subsection{Implications of Findings}

\subsubsection{Implications for Researchers}
Our research demonstrates that current LCMs perform variously in the three scenarios of the API suggestion task. With well-designed context selection, the model performance can be substantially improved. However, as shown in RQ1 and RQ2, there exists a remarkable gap between the performance in the ``when to use'' and the other scenarios, indicating the need for further research and improvement in this direction.

Our results also reveal the potential research directions in the era of LCM for the community. Specifically:
\begin{itemize}
    \item \textbf{Exploring more effective context selection approaches}. This paper concentrates on selecting contexts within the code file. As reported in the work \cite{nashid2024contextual, shrivastava2023repository}, cross-file information can potentially enhance API suggestions. Therefore, it is essential to explore more effective context selection approaches that reduce token length while simultaneously improving model performance. Future research should investigate methods such as analyzing file dependencies and selecting context from other related files within the project.

    \item \textbf{Paying more attention to the ``when to use'' scenarios}. Previous research has predominantly concentrated on suggesting which APIs to use, ignoring other scenarios and challenges developers encounter when working with APIs. Our results in RQ1 and RQ2 demonstrate that LCMs exhibit varying performance across different scenarios. Notably, in the ``when to use'' scenario, we can find that the average exact match metric is 16\% and 22.4\% lower than that in the ``which to use'' and ``how to use'' scenarios (with the optimal context selection), respectively.  Therefore, these findings suggest that researchers should broaden their focus to include a wider range of scenarios in API suggestion tasks, better improving productivity for developers.
\end{itemize}

\subsubsection{Implications for Developers}
In this section, we take both performance and efficiency into account and provide insights on model selection and context selection.

\begin{figure}
    \centering
    \includegraphics[width=0.49\textwidth]{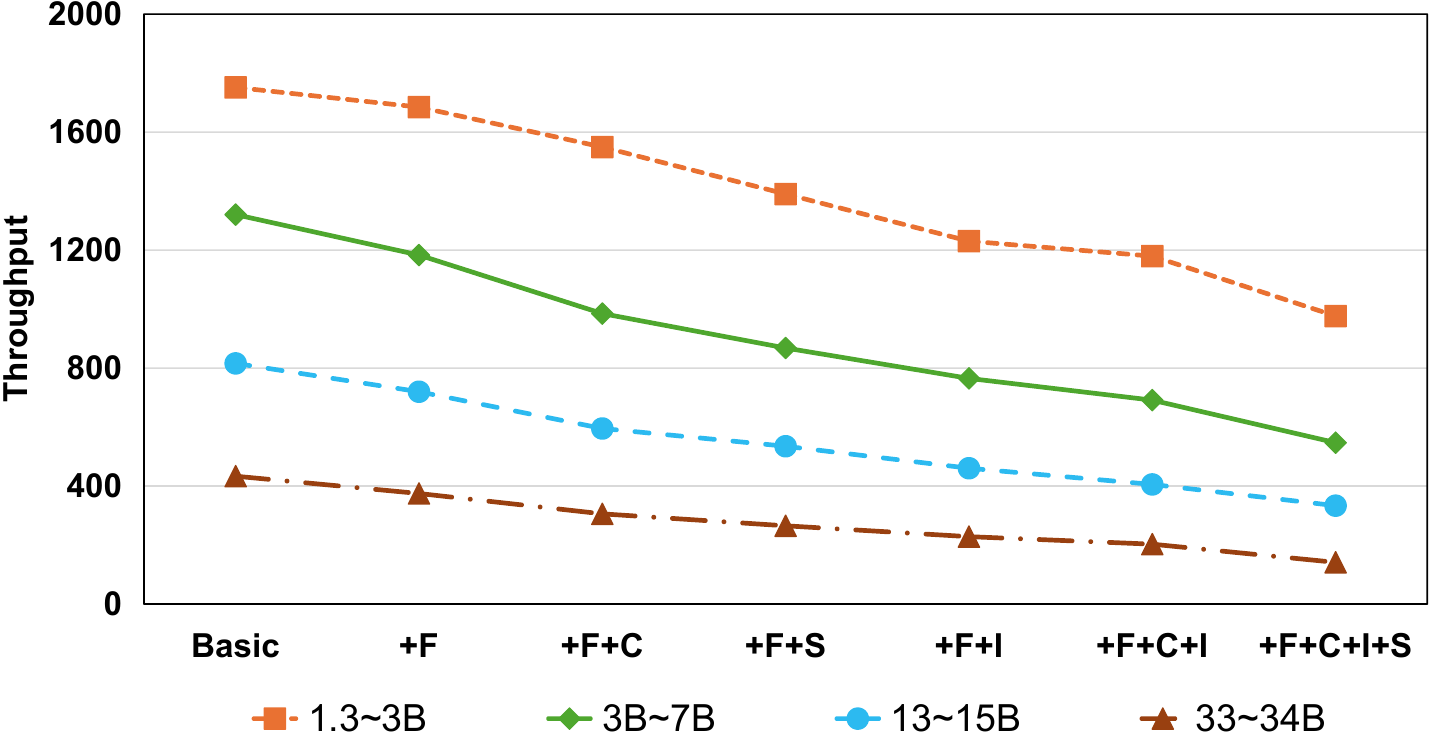}
    \vspace{-8pt}
    \caption{Illustration of throughput of the different contexts, in which the models are grouped by their sizes for clarity.}
    \label{fig:scenario_and_throughput}
    \vspace{-12pt}
\end{figure}

\textbf{Implications on model selection}. In Section \ref{sec:rq2}, we establish that larger models generally outperform smaller ones and report the average throughput of nine LCMs with the provided contexts. To further inform model selection, we present the throughput of LCMs that are grouped according to their sizes in Figure \ref{fig:scenario_and_throughput} (the detailed results of individual LCMs are involved in our anonymous repository). The figure reveals that all experimented LCMs exhibit similar trends across different context combinations. Notably, we observe that smaller LCMs (e.g., StarCoder 3B and DeepSeek-Coder 1.3B) maintain higher throughput across all contexts (+F+C+I+S) compared to larger LCMs (e.g., StarCoder 15B and DeepSeek-Coder 33B). These findings suggest that if computational resources cannot accommodate large LCMs with over 10 billion parameters, opting for smaller models with enriched contexts is a more efficient and effective choice in the API suggestion task.  

In addition, the combination of different LCMs according to scenarios is also worth exploring.
As demonstrated in Section \ref{sec:rq2}, in simpler scenarios such as the ``how to use'' scenario, the performance gap between large models and small models is relatively small. 
Therefore, for less demanding scenarios such as ``how to use'', smaller models can be utilized to achieve satisfactory performance while conserving computational resources and reducing latency. Conversely, for more complex scenarios that require higher accuracy and nuanced understanding, larger models can be employed to leverage their superior capabilities. This approach allows for efficient resource management and optimized performance across a variety of use cases.

\textbf{Implications on context selection.}
According to Figure \ref{fig:efficiency}, we observe that different contexts contribute variously across different scenarios, indicating that there is no universal solution for context selection in API suggestion tasks. For instance, combining all available contexts (+F+C+I+S) yields the best performance in the three scenarios. However, this improvement is less satisfactory when considering the 54\% throughput overhead in the ``how to use'' scenario.

Therefore, when deploying API suggestion services with limited computational resources that cannot accommodate all studied contexts, developers could tailor context selection to specific scenarios. If the latency associated with loading all contexts (+F+C+I+S) is unacceptable, using comments (+C) and suffix contexts (+S) can be effective choices for the "how to use" and "which to use" scenarios. For the "when to use" scenario, combining comments and import messages (+C+I) offers a good trade-off between performance and computational efficiency.


\subsection{Discussion about Evaluation Metrics}
{In this paper, we use edit similarity to evaluate the string-level similarity of LCMs' outputs and ground truth API calls. Specifically, we use the metric due to the following reasons:}
\begin{itemize}
    \item {API suggestion is a specific scenario in code completion tasks and edit similarity is a popular metric used in existing research \cite{lu2021codexglue}.}

    \item {The edit similarity metric is meaningful in API suggestion tasks. For instance, LCMs may predict a wrong but similar API to use, e.g., the target is getField but predicting getFields. In this situation, the prediction has a high edit similarity, indicating that users just need to delete an ``s'' after accepting LCMs' output. Therefore, edit similarity can help identify a ``plausible'' answer that is close to the users' needs.}

    \item {The ranking metric is also helpful; however, calculating this metric requires LCMs to generate multiple answers, which brings substantial overhead on time consumption. We will involve this metric in future work.}
\end{itemize}

\subsection{Threats to Validity}
We have identified the following major threats to validity:

\textbf{Limited LCMs}. The experiments in this paper are based on open-source popular LCMs, which may bring bias in the results. To mitigate this issue, we select nine LCMs with various model sizes to control the threat. Furthermore, the improvement of context selection is model-agnostic, making our findings easily generalize to other LCMs.

\textbf{Limited API source}. In this paper, we focus on the Spring Framework in Java to evaluate the API suggestion task. This choice is grounded that SpringFramework is one of the most widely used frameworks in the Java ecosystem, offering a comprehensive set of features for building robust and scalable applications. In addition, the diversity and complexity of SpringFramework APIs provide a challenging testbed for evaluating the performance of large code models (LCMs) in API suggestion tasks.

{\textbf{Potential data leakage.} In this paper, the data used for training LCMs
are not publicly available, so that
we can hardly determine whether there exists data leakage in these models. However, our experiments reveal that merely providing function contexts to LCMs cannot yield promising performance. Therefore, we believe that the results produced by LCMs in the benchmark are not from simply memorizing data.}

\section{Related work}
\subsection{API Suggestion}
Application Programming Interfaces (APIs) are crucial for enabling developers to integrate existing functionalities rather than building them from scratch. Various automated API method recommendation techniques have been developed to assist developers in writing correct APIs~\cite{rahman2016rack,zhou2021boosting,DBLP:conf/icse/McMillanGPXF11}. McMillan et al.~\cite{DBLP:conf/icse/McMillanGPXF11} propose portfolio, an API recommendation tool that aids programmers in locating relevant code snippets that fulfill high-level requirements specified in a query. Gu et al.~\cite{DBLP:conf/sigsoft/GuZZK16} propose DeepAPI, which utilizes a deep learning model to suggest API usage sequences for a given natural language query. CLEAR~\cite{wei2022clear} uses contrastive learning and  BERT sentence embedding similarity to first identify a set of candidate Stack Overflow posts and then re-ranks to recommend the top-N APIs. MEGA~\cite{DBLP:journals/tse/ChenGRP0L23} employs heterogeneous graphs to learn the matching scores between methods and APIs for recommending related APIs. Different from previous works that mainly focus on API recommendation, our work encompasses more scenarios in API suggestion including when, which, and how to use. 

\subsection{Large Code Models}
Recently, the emergence of Large code models has revolutionized various software engineering tasks~\cite{roziere2023code,gao2024learning,DBLP:journals/corr/abs-2312-02120,fan2023large}.  StarCoder is a foundation code model~\cite{lozhkov2024starcoder} trained on the mixture of source code and natural language texts. Its training data incorporate more than 80 different programming languages as well as text extracted from GitHub issues and commits and from notebooks. Code Llama~\cite{roziere2023code} is a foundation model developed by Meta that helps generate and understand programming code. It builds on the capabilities of the original LLaMA models and extends the context length to 16K. DeepSeek Coder~\cite{guo2024deepseek} has a range of open-source LCMs with sizes varying from 1.3B to 33B. It is trained from scratch on a curated code corpus with 2T tokens. Apart from the foundation models, various fine-tuning~\cite{DBLP:journals/corr/abs-2312-02120,luo2023wizardcoder, wang2022no} and prompting techniques~\cite{DBLP:conf/icse/AhmedPDB24,gao2023makes} are also proposed to make full use of LCMs for software engineering tasks. For example, Magicoder~\cite{DBLP:journals/corr/abs-2312-02120} proposes to synthesize instruction data from open-source code snippets for effectively tuning large code models for code generation. Ahmed et al.~\cite{DBLP:conf/icse/AhmedPDB24} propose to augment the prompt with repository information and data flow for boosting the performance of code summarization. TypeGEN~\cite{DBLP:conf/kbse/PengWWGL23} uses static analysis results and chain-of-thought prompts to guide LLMs in type inference. ChatUniTest~\cite{DBLP:journals/corr/abs-2305-04764} extracts essential information and creates an adaptive focal context for LCMs to generate test cases.

\subsection{LLMs in API Suggestion}
{Huang et al. \cite{huang2023let} propose to combine knowledge graphs and LLMs to find APIs based on natural language queries. CAPIR \cite{ma2024compositional} adopts a ``divide-and-conquer'' strategy to recommend APIs for coarse-grained developmental requirements. APIGen \cite{chen2024apigen} is a generative method that utilizes improved in-context learning to directly generate the API name. Recently, Nashid et al. \cite{nashid2024contextual} study the performance of LLMs in generating APIs for unseen repositories. Different from previous works that only focus on API recommendation (i.e., which to use), our work presents a more comprehensive study of different API usage scenarios in real-world software development. }
\section{Conclusion}
In this paper, we have conducted a systematic evaluation of LCMs in API suggestion, proposing three distinct scenarios, when, which, and how to use an API, to comprehensively assess their capabilities. Our experiments on a diverse benchmark dataset have revealed that LCMs perform best in the ``how to use'' scenario and benefit substantially from enriched context, albeit at the cost of increased token length and reduced throughput.
Our findings offer valuable insights into the strengths and limitations of LCMs in API suggestion and highlight the critical role of context in enhancing their performance.


\bibliographystyle{ACM-Reference-Format}
\bibliography{acmart}


\begin{thebibliography}{44}


\ifx \showCODEN    \undefined \def \showCODEN     #1{\unskip}     \fi
\ifx \showDOI      \undefined \def \showDOI       #1{#1}\fi
\ifx \showISBNx    \undefined \def \showISBNx     #1{\unskip}     \fi
\ifx \showISBNxiii \undefined \def \showISBNxiii  #1{\unskip}     \fi
\ifx \showISSN     \undefined \def \showISSN      #1{\unskip}     \fi
\ifx \showLCCN     \undefined \def \showLCCN      #1{\unskip}     \fi
\ifx \shownote     \undefined \def \shownote      #1{#1}          \fi
\ifx \showarticletitle \undefined \def \showarticletitle #1{#1}   \fi
\ifx \showURL      \undefined \def \showURL       {\relax}        \fi
\providecommand\bibfield[2]{#2}
\providecommand\bibinfo[2]{#2}
\providecommand\natexlab[1]{#1}
\providecommand\showeprint[2][]{arXiv:#2}

\bibitem[mav({[n.\,d.]})]%
        {maven}
 \bibinfo{year}{[n.\,d.]}\natexlab{}.
\newblock \bibinfo{booktitle}{\emph{{MVN Repositories}}}.
\newblock
\newblock
\shownote{\url{https://mvnrepository.com/open-source/web-frameworks}}.


\bibitem[Agrawal et~al\mbox{.}(2024)]%
        {agrawal2024taming}
\bibfield{author}{\bibinfo{person}{Amey Agrawal}, \bibinfo{person}{Nitin Kedia}, \bibinfo{person}{Ashish Panwar}, \bibinfo{person}{Jayashree Mohan}, \bibinfo{person}{Nipun Kwatra}, \bibinfo{person}{Bhargav~S Gulavani}, \bibinfo{person}{Alexey Tumanov}, {and} \bibinfo{person}{Ramachandran Ramjee}.} \bibinfo{year}{2024}\natexlab{}.
\newblock \showarticletitle{Taming Throughput-Latency Tradeoff in LLM Inference with Sarathi-Serve}.
\newblock \bibinfo{journal}{\emph{arXiv preprint arXiv:2403.02310}} (\bibinfo{year}{2024}).
\newblock


\bibitem[Ahmed et~al\mbox{.}(2024)]%
        {DBLP:conf/icse/AhmedPDB24}
\bibfield{author}{\bibinfo{person}{Toufique Ahmed}, \bibinfo{person}{Kunal~Suresh Pai}, \bibinfo{person}{Premkumar~T. Devanbu}, {and} \bibinfo{person}{Earl~T. Barr}.} \bibinfo{year}{2024}\natexlab{}.
\newblock \showarticletitle{Automatic Semantic Augmentation of Language Model Prompts (for Code Summarization)}. In \bibinfo{booktitle}{\emph{Proceedings of the 46th {IEEE/ACM} International Conference on Software Engineering, {ICSE} 2024, Lisbon, Portugal, April 14-20, 2024}}. \bibinfo{publisher}{{ACM}}, \bibinfo{pages}{220:1--220:13}.
\newblock


\bibitem[Brown et~al\mbox{.}(2020)]%
        {Brown2020LanguageMA}
\bibfield{author}{\bibinfo{person}{Tom~B. Brown}, \bibinfo{person}{Benjamin Mann}, \bibinfo{person}{Nick Ryder}, \bibinfo{person}{Melanie Subbiah}, \bibinfo{person}{Jared Kaplan}, \bibinfo{person}{Prafulla Dhariwal}, \bibinfo{person}{Arvind Neelakantan}, \bibinfo{person}{Pranav Shyam}, \bibinfo{person}{Girish Sastry}, \bibinfo{person}{Amanda Askell}, \bibinfo{person}{Sandhini Agarwal}, \bibinfo{person}{Ariel Herbert{-}Voss}, \bibinfo{person}{Gretchen Krueger}, \bibinfo{person}{Tom Henighan}, \bibinfo{person}{Rewon Child}, \bibinfo{person}{Aditya Ramesh}, \bibinfo{person}{Daniel~M. Ziegler}, \bibinfo{person}{Jeffrey Wu}, \bibinfo{person}{Clemens Winter}, \bibinfo{person}{Christopher Hesse}, \bibinfo{person}{Mark Chen}, \bibinfo{person}{Eric Sigler}, \bibinfo{person}{Mateusz Litwin}, \bibinfo{person}{Scott Gray}, \bibinfo{person}{Benjamin Chess}, \bibinfo{person}{Jack Clark}, \bibinfo{person}{Christopher Berner}, \bibinfo{person}{Sam McCandlish}, \bibinfo{person}{Alec Radford}, \bibinfo{person}{Ilya Sutskever},
  {and} \bibinfo{person}{Dario Amodei}.} \bibinfo{year}{2020}\natexlab{}.
\newblock \showarticletitle{Language Models are Few-Shot Learners}.
\newblock  (\bibinfo{year}{2020}).
\newblock


\bibitem[Chen et~al\mbox{.}(2023)]%
        {DBLP:journals/tse/ChenGRP0L23}
\bibfield{author}{\bibinfo{person}{Yujia Chen}, \bibinfo{person}{Cuiyun Gao}, \bibinfo{person}{Xiaoxue Ren}, \bibinfo{person}{Yun Peng}, \bibinfo{person}{Xin Xia}, {and} \bibinfo{person}{Michael~R. Lyu}.} \bibinfo{year}{2023}\natexlab{}.
\newblock \showarticletitle{{API} Usage Recommendation Via Multi-View Heterogeneous Graph Representation Learning}.
\newblock \bibinfo{journal}{\emph{{IEEE} Trans. Software Eng.}} \bibinfo{volume}{49}, \bibinfo{number}{5} (\bibinfo{year}{2023}), \bibinfo{pages}{3289--3304}.
\newblock


\bibitem[Chen et~al\mbox{.}(2024)]%
        {chen2024apigen}
\bibfield{author}{\bibinfo{person}{Yujia Chen}, \bibinfo{person}{Cuiyun Gao}, \bibinfo{person}{Muyijie Zhu}, \bibinfo{person}{Qing Liao}, \bibinfo{person}{Yong Wang}, {and} \bibinfo{person}{Guoai Xu}.} \bibinfo{year}{2024}\natexlab{}.
\newblock \showarticletitle{APIGen: Generative API Method Recommendation}.
\newblock \bibinfo{journal}{\emph{arXiv preprint arXiv:2401.15843}} (\bibinfo{year}{2024}).
\newblock


\bibitem[Ciniselli et~al\mbox{.}(2021)]%
        {ciniselli2021empirical}
\bibfield{author}{\bibinfo{person}{Matteo Ciniselli}, \bibinfo{person}{Nathan Cooper}, \bibinfo{person}{Luca Pascarella}, \bibinfo{person}{Denys Poshyvanyk}, \bibinfo{person}{Massimiliano Di~Penta}, {and} \bibinfo{person}{Gabriele Bavota}.} \bibinfo{year}{2021}\natexlab{}.
\newblock \showarticletitle{An empirical study on the usage of bert models for code completion}. In \bibinfo{booktitle}{\emph{2021 IEEE/ACM 18th International Conference on Mining Software Repositories (MSR)}}. IEEE, \bibinfo{pages}{108--119}.
\newblock


\bibitem[Dao et~al\mbox{.}(2022)]%
        {dao2022flashattention}
\bibfield{author}{\bibinfo{person}{Tri Dao}, \bibinfo{person}{Dan Fu}, \bibinfo{person}{Stefano Ermon}, \bibinfo{person}{Atri Rudra}, {and} \bibinfo{person}{Christopher R{\'e}}.} \bibinfo{year}{2022}\natexlab{}.
\newblock \showarticletitle{Flashattention: Fast and memory-efficient exact attention with io-awareness}.
\newblock \bibinfo{journal}{\emph{Advances in Neural Information Processing Systems}}  \bibinfo{volume}{35} (\bibinfo{year}{2022}), \bibinfo{pages}{16344--16359}.
\newblock


\bibitem[Fan et~al\mbox{.}(2023)]%
        {fan2023large}
\bibfield{author}{\bibinfo{person}{Angela Fan}, \bibinfo{person}{Beliz Gokkaya}, \bibinfo{person}{Mark Harman}, \bibinfo{person}{Mitya Lyubarskiy}, \bibinfo{person}{Shubho Sengupta}, \bibinfo{person}{Shin Yoo}, {and} \bibinfo{person}{Jie~M Zhang}.} \bibinfo{year}{2023}\natexlab{}.
\newblock \showarticletitle{Large language models for software engineering: Survey and open problems}.
\newblock \bibinfo{journal}{\emph{arXiv preprint arXiv:2310.03533}} (\bibinfo{year}{2023}).
\newblock


\bibitem[Fowkes and Sutton(2016)]%
        {fowkes2016parameter}
\bibfield{author}{\bibinfo{person}{Jaroslav Fowkes} {and} \bibinfo{person}{Charles Sutton}.} \bibinfo{year}{2016}\natexlab{}.
\newblock \showarticletitle{Parameter-free probabilistic API mining across GitHub}. In \bibinfo{booktitle}{\emph{Proceedings of the 2016 24th ACM SIGSOFT international symposium on foundations of software engineering}}. \bibinfo{pages}{254--265}.
\newblock


\bibitem[Gao et~al\mbox{.}(2023a)]%
        {DBLP:journals/tosem/GaoGHZNXL23}
\bibfield{author}{\bibinfo{person}{Shuzheng Gao}, \bibinfo{person}{Cuiyun Gao}, \bibinfo{person}{Yulan He}, \bibinfo{person}{Jichuan Zeng}, \bibinfo{person}{Lunyiu Nie}, \bibinfo{person}{Xin Xia}, {and} \bibinfo{person}{Michael~R. Lyu}.} \bibinfo{year}{2023}\natexlab{a}.
\newblock \showarticletitle{Code Structure-Guided Transformer for Source Code Summarization}.
\newblock \bibinfo{journal}{\emph{{ACM} Trans. Softw. Eng. Methodol.}} \bibinfo{volume}{32}, \bibinfo{number}{1} (\bibinfo{year}{2023}), \bibinfo{pages}{23:1--23:32}.
\newblock


\bibitem[Gao et~al\mbox{.}(2024)]%
        {gao2024learning}
\bibfield{author}{\bibinfo{person}{Shuzheng Gao}, \bibinfo{person}{Wenxin Mao}, \bibinfo{person}{Cuiyun Gao}, \bibinfo{person}{Li Li}, \bibinfo{person}{Xing Hu}, \bibinfo{person}{Xin Xia}, {and} \bibinfo{person}{Michael~R Lyu}.} \bibinfo{year}{2024}\natexlab{}.
\newblock \showarticletitle{Learning in the wild: Towards leveraging unlabeled data for effectively tuning pre-trained code models}. In \bibinfo{booktitle}{\emph{Proceedings of the IEEE/ACM 46th International Conference on Software Engineering}}. \bibinfo{pages}{1--13}.
\newblock


\bibitem[Gao et~al\mbox{.}(2023b)]%
        {gao2023makes}
\bibfield{author}{\bibinfo{person}{Shuzheng Gao}, \bibinfo{person}{Xin-Cheng Wen}, \bibinfo{person}{Cuiyun Gao}, \bibinfo{person}{Wenxuan Wang}, \bibinfo{person}{Hongyu Zhang}, {and} \bibinfo{person}{Michael~R Lyu}.} \bibinfo{year}{2023}\natexlab{b}.
\newblock \showarticletitle{What makes good in-context demonstrations for code intelligence tasks with llms?}. In \bibinfo{booktitle}{\emph{2023 38th IEEE/ACM International Conference on Automated Software Engineering (ASE)}}. IEEE, \bibinfo{pages}{761--773}.
\newblock


\bibitem[Garousi et~al\mbox{.}(2015)]%
        {garousi2015usage}
\bibfield{author}{\bibinfo{person}{Golara Garousi}, \bibinfo{person}{Vahid Garousi-Yusifo{\u{g}}lu}, \bibinfo{person}{Guenther Ruhe}, \bibinfo{person}{Junji Zhi}, \bibinfo{person}{Mahmoud Moussavi}, {and} \bibinfo{person}{Brian Smith}.} \bibinfo{year}{2015}\natexlab{}.
\newblock \showarticletitle{Usage and usefulness of technical software documentation: An industrial case study}.
\newblock \bibinfo{journal}{\emph{Information and software technology}}  \bibinfo{volume}{57} (\bibinfo{year}{2015}), \bibinfo{pages}{664--682}.
\newblock


\bibitem[Gu et~al\mbox{.}(2016)]%
        {DBLP:conf/sigsoft/GuZZK16}
\bibfield{author}{\bibinfo{person}{Xiaodong Gu}, \bibinfo{person}{Hongyu Zhang}, \bibinfo{person}{Dongmei Zhang}, {and} \bibinfo{person}{Sunghun Kim}.} \bibinfo{year}{2016}\natexlab{}.
\newblock \showarticletitle{Deep {API} learning}. In \bibinfo{booktitle}{\emph{Proceedings of the 24th {ACM} {SIGSOFT} International Symposium on Foundations of Software Engineering, {FSE} 2016, Seattle, WA, USA, November 13-18, 2016}}. \bibinfo{publisher}{{ACM}}, \bibinfo{pages}{631--642}.
\newblock


\bibitem[Guo et~al\mbox{.}(2024)]%
        {guo2024deepseek}
\bibfield{author}{\bibinfo{person}{Daya Guo}, \bibinfo{person}{Qihao Zhu}, \bibinfo{person}{Dejian Yang}, \bibinfo{person}{Zhenda Xie}, \bibinfo{person}{Kai Dong}, \bibinfo{person}{Wentao Zhang}, \bibinfo{person}{Guanting Chen}, \bibinfo{person}{Xiao Bi}, \bibinfo{person}{Y. Wu}, \bibinfo{person}{Y.~K. Li}, \bibinfo{person}{Fuli Luo}, \bibinfo{person}{Yingfei Xiong}, {and} \bibinfo{person}{Wenfeng Liang}.} \bibinfo{year}{2024}\natexlab{}.
\newblock \showarticletitle{DeepSeek-Coder: When the Large Language Model Meets Programming - The Rise of Code Intelligence}.
\newblock \bibinfo{journal}{\emph{CoRR}}  \bibinfo{volume}{abs/2401.14196} (\bibinfo{year}{2024}).
\newblock


\bibitem[He et~al\mbox{.}(2021)]%
        {he2021pyart}
\bibfield{author}{\bibinfo{person}{Xincheng He}, \bibinfo{person}{Lei Xu}, \bibinfo{person}{Xiangyu Zhang}, \bibinfo{person}{Rui Hao}, \bibinfo{person}{Yang Feng}, {and} \bibinfo{person}{Baowen Xu}.} \bibinfo{year}{2021}\natexlab{}.
\newblock \showarticletitle{Pyart: Python api recommendation in real-time}. In \bibinfo{booktitle}{\emph{2021 IEEE/ACM 43rd International Conference on Software Engineering (ICSE)}}. IEEE, \bibinfo{pages}{1634--1645}.
\newblock


\bibitem[Huang et~al\mbox{.}(2023)]%
        {huang2023let}
\bibfield{author}{\bibinfo{person}{Qing Huang}, \bibinfo{person}{Zhenyu Wan}, \bibinfo{person}{Zhenchang Xing}, \bibinfo{person}{Changjing Wang}, \bibinfo{person}{Jieshan Chen}, \bibinfo{person}{Xiwei Xu}, {and} \bibinfo{person}{Qinghua Lu}.} \bibinfo{year}{2023}\natexlab{}.
\newblock \showarticletitle{Let's Chat to Find the APIs: Connecting Human, LLM and Knowledge Graph through AI Chain}. In \bibinfo{booktitle}{\emph{2023 38th IEEE/ACM International Conference on Automated Software Engineering (ASE)}}. IEEE, \bibinfo{pages}{471--483}.
\newblock


\bibitem[Kwon et~al\mbox{.}(2023)]%
        {kwon2023efficient}
\bibfield{author}{\bibinfo{person}{Woosuk Kwon}, \bibinfo{person}{Zhuohan Li}, \bibinfo{person}{Siyuan Zhuang}, \bibinfo{person}{Ying Sheng}, \bibinfo{person}{Lianmin Zheng}, \bibinfo{person}{Cody~Hao Yu}, \bibinfo{person}{Joseph Gonzalez}, \bibinfo{person}{Hao Zhang}, {and} \bibinfo{person}{Ion Stoica}.} \bibinfo{year}{2023}\natexlab{}.
\newblock \showarticletitle{Efficient memory management for large language model serving with pagedattention}. In \bibinfo{booktitle}{\emph{Proceedings of the 29th Symposium on Operating Systems Principles}}. \bibinfo{pages}{611--626}.
\newblock


\bibitem[Li et~al\mbox{.}(2023)]%
        {li2023starcoder}
\bibfield{author}{\bibinfo{person}{Raymond Li}, \bibinfo{person}{Loubna~Ben Allal}, \bibinfo{person}{Yangtian Zi}, \bibinfo{person}{Niklas Muennighoff}, \bibinfo{person}{Denis Kocetkov}, \bibinfo{person}{Chenghao Mou}, \bibinfo{person}{Marc Marone}, \bibinfo{person}{Christopher Akiki}, \bibinfo{person}{Jia Li}, \bibinfo{person}{Jenny Chim}, \bibinfo{person}{Qian Liu}, \bibinfo{person}{Evgenii Zheltonozhskii}, \bibinfo{person}{Terry~Yue Zhuo}, \bibinfo{person}{Thomas Wang}, \bibinfo{person}{Olivier Dehaene}, \bibinfo{person}{Mishig Davaadorj}, \bibinfo{person}{Joel Lamy{-}Poirier}, \bibinfo{person}{Jo{\~{a}}o Monteiro}, \bibinfo{person}{Oleh Shliazhko}, \bibinfo{person}{Nicolas Gontier}, \bibinfo{person}{Nicholas Meade}, \bibinfo{person}{Armel Zebaze}, \bibinfo{person}{Ming{-}Ho Yee}, \bibinfo{person}{Logesh~Kumar Umapathi}, \bibinfo{person}{Jian Zhu}, \bibinfo{person}{Benjamin Lipkin}, \bibinfo{person}{Muhtasham Oblokulov}, \bibinfo{person}{Zhiruo Wang}, \bibinfo{person}{Rudra~Murthy V},
  \bibinfo{person}{Jason Stillerman}, \bibinfo{person}{Siva~Sankalp Patel}, \bibinfo{person}{Dmitry Abulkhanov}, \bibinfo{person}{Marco Zocca}, \bibinfo{person}{Manan Dey}, \bibinfo{person}{Zhihan Zhang}, \bibinfo{person}{Nour Moustafa{-}Fahmy}, \bibinfo{person}{Urvashi Bhattacharyya}, \bibinfo{person}{Wenhao Yu}, \bibinfo{person}{Swayam Singh}, \bibinfo{person}{Sasha Luccioni}, \bibinfo{person}{Paulo Villegas}, \bibinfo{person}{Maxim Kunakov}, \bibinfo{person}{Fedor Zhdanov}, \bibinfo{person}{Manuel Romero}, \bibinfo{person}{Tony Lee}, \bibinfo{person}{Nadav Timor}, \bibinfo{person}{Jennifer Ding}, \bibinfo{person}{Claire Schlesinger}, \bibinfo{person}{Hailey Schoelkopf}, \bibinfo{person}{Jan Ebert}, \bibinfo{person}{Tri Dao}, \bibinfo{person}{Mayank Mishra}, \bibinfo{person}{Alex Gu}, \bibinfo{person}{Jennifer Robinson}, \bibinfo{person}{Carolyn~Jane Anderson}, \bibinfo{person}{Brendan Dolan{-}Gavitt}, \bibinfo{person}{Danish Contractor}, \bibinfo{person}{Siva Reddy}, \bibinfo{person}{Daniel Fried},
  \bibinfo{person}{Dzmitry Bahdanau}, \bibinfo{person}{Yacine Jernite}, \bibinfo{person}{Carlos~Mu{\~{n}}oz Ferrandis}, \bibinfo{person}{Sean Hughes}, \bibinfo{person}{Thomas Wolf}, \bibinfo{person}{Arjun Guha}, \bibinfo{person}{Leandro von Werra}, {and} \bibinfo{person}{Harm de Vries}.} \bibinfo{year}{2023}\natexlab{}.
\newblock \showarticletitle{StarCoder: may the source be with you!}
\newblock \bibinfo{journal}{\emph{CoRR}}  \bibinfo{volume}{abs/2305.06161} (\bibinfo{year}{2023}).
\newblock


\bibitem[Liu et~al\mbox{.}(2024)]%
        {liu2024non}
\bibfield{author}{\bibinfo{person}{Fang Liu}, \bibinfo{person}{Zhiyi Fu}, \bibinfo{person}{Ge Li}, \bibinfo{person}{Zhi Jin}, \bibinfo{person}{Hui Liu}, \bibinfo{person}{Yiyang Hao}, {and} \bibinfo{person}{Li Zhang}.} \bibinfo{year}{2024}\natexlab{}.
\newblock \showarticletitle{Non-Autoregressive Line-Level Code Completion}.
\newblock \bibinfo{journal}{\emph{ACM Transactions on Software Engineering and Methodology}} (\bibinfo{year}{2024}).
\newblock


\bibitem[Lozhkov et~al\mbox{.}(2024)]%
        {lozhkov2024starcoder}
\bibfield{author}{\bibinfo{person}{Anton Lozhkov}, \bibinfo{person}{Raymond Li}, \bibinfo{person}{Loubna~Ben Allal}, \bibinfo{person}{Federico Cassano}, \bibinfo{person}{Joel Lamy{-}Poirier}, \bibinfo{person}{Nouamane Tazi}, \bibinfo{person}{Ao Tang}, \bibinfo{person}{Dmytro Pykhtar}, \bibinfo{person}{Jiawei Liu}, \bibinfo{person}{Yuxiang Wei}, \bibinfo{person}{Tianyang Liu}, \bibinfo{person}{Max Tian}, \bibinfo{person}{Denis Kocetkov}, \bibinfo{person}{Arthur Zucker}, \bibinfo{person}{Younes Belkada}, \bibinfo{person}{Zijian Wang}, \bibinfo{person}{Qian Liu}, \bibinfo{person}{Dmitry Abulkhanov}, \bibinfo{person}{Indraneil Paul}, \bibinfo{person}{Zhuang Li}, \bibinfo{person}{Wen{-}Ding Li}, \bibinfo{person}{Megan Risdal}, \bibinfo{person}{Jia Li}, \bibinfo{person}{Jian Zhu}, \bibinfo{person}{Terry~Yue Zhuo}, \bibinfo{person}{Evgenii Zheltonozhskii}, \bibinfo{person}{Nii Osae~Osae Dade}, \bibinfo{person}{Wenhao Yu}, \bibinfo{person}{Lucas Krau{\ss}}, \bibinfo{person}{Naman Jain}, \bibinfo{person}{Yixuan Su},
  \bibinfo{person}{Xuanli He}, \bibinfo{person}{Manan Dey}, \bibinfo{person}{Edoardo Abati}, \bibinfo{person}{Yekun Chai}, \bibinfo{person}{Niklas Muennighoff}, \bibinfo{person}{Xiangru Tang}, \bibinfo{person}{Muhtasham Oblokulov}, \bibinfo{person}{Christopher Akiki}, \bibinfo{person}{Marc Marone}, \bibinfo{person}{Chenghao Mou}, \bibinfo{person}{Mayank Mishra}, \bibinfo{person}{Alex Gu}, \bibinfo{person}{Binyuan Hui}, \bibinfo{person}{Tri Dao}, \bibinfo{person}{Armel Zebaze}, \bibinfo{person}{Olivier Dehaene}, \bibinfo{person}{Nicolas Patry}, \bibinfo{person}{Canwen Xu}, \bibinfo{person}{Julian~J. McAuley}, \bibinfo{person}{Han Hu}, \bibinfo{person}{Torsten Scholak}, \bibinfo{person}{S{\'{e}}bastien Paquet}, \bibinfo{person}{Jennifer Robinson}, \bibinfo{person}{Carolyn~Jane Anderson}, \bibinfo{person}{Nicolas Chapados}, {and} \bibinfo{person}{et al.}} \bibinfo{year}{2024}\natexlab{}.
\newblock \showarticletitle{StarCoder 2 and The Stack v2: The Next Generation}.
\newblock \bibinfo{journal}{\emph{CoRR}}  \bibinfo{volume}{abs/2402.19173} (\bibinfo{year}{2024}).
\newblock


\bibitem[Lu et~al\mbox{.}(2021)]%
        {lu2021codexglue}
\bibfield{author}{\bibinfo{person}{Shuai Lu}, \bibinfo{person}{Daya Guo}, \bibinfo{person}{Shuo Ren}, \bibinfo{person}{Junjie Huang}, \bibinfo{person}{Alexey Svyatkovskiy}, \bibinfo{person}{Ambrosio Blanco}, \bibinfo{person}{Colin~B. Clement}, \bibinfo{person}{Dawn Drain}, \bibinfo{person}{Daxin Jiang}, \bibinfo{person}{Duyu Tang}, \bibinfo{person}{Ge Li}, \bibinfo{person}{Lidong Zhou}, \bibinfo{person}{Linjun Shou}, \bibinfo{person}{Long Zhou}, \bibinfo{person}{Michele Tufano}, \bibinfo{person}{Ming Gong}, \bibinfo{person}{Ming Zhou}, \bibinfo{person}{Nan Duan}, \bibinfo{person}{Neel Sundaresan}, \bibinfo{person}{Shao~Kun Deng}, \bibinfo{person}{Shengyu Fu}, {and} \bibinfo{person}{Shujie Liu}.} \bibinfo{year}{2021}\natexlab{}.
\newblock \showarticletitle{CodeXGLUE: {A} Machine Learning Benchmark Dataset for Code Understanding and Generation}.
\newblock  (\bibinfo{year}{2021}).
\newblock


\bibitem[Luo et~al\mbox{.}(2023)]%
        {luo2023wizardcoder}
\bibfield{author}{\bibinfo{person}{Ziyang Luo}, \bibinfo{person}{Can Xu}, \bibinfo{person}{Pu Zhao}, \bibinfo{person}{Qingfeng Sun}, \bibinfo{person}{Xiubo Geng}, \bibinfo{person}{Wenxiang Hu}, \bibinfo{person}{Chongyang Tao}, \bibinfo{person}{Jing Ma}, \bibinfo{person}{Qingwei Lin}, {and} \bibinfo{person}{Daxin Jiang}.} \bibinfo{year}{2023}\natexlab{}.
\newblock \showarticletitle{Wizardcoder: Empowering code large language models with evol-instruct}.
\newblock \bibinfo{journal}{\emph{arXiv preprint arXiv:2306.08568}} (\bibinfo{year}{2023}).
\newblock


\bibitem[Ma et~al\mbox{.}(2024)]%
        {ma2024compositional}
\bibfield{author}{\bibinfo{person}{Zexiong Ma}, \bibinfo{person}{Shengnan An}, \bibinfo{person}{Bing Xie}, {and} \bibinfo{person}{Zeqi Lin}.} \bibinfo{year}{2024}\natexlab{}.
\newblock \showarticletitle{Compositional API Recommendation for Library-Oriented Code Generation}. In \bibinfo{booktitle}{\emph{Proceedings of the 32nd IEEE/ACM International Conference on Program Comprehension}}. \bibinfo{pages}{87--98}.
\newblock


\bibitem[McMillan et~al\mbox{.}(2011)]%
        {DBLP:conf/icse/McMillanGPXF11}
\bibfield{author}{\bibinfo{person}{Collin McMillan}, \bibinfo{person}{Mark Grechanik}, \bibinfo{person}{Denys Poshyvanyk}, \bibinfo{person}{Qing Xie}, {and} \bibinfo{person}{Chen Fu}.} \bibinfo{year}{2011}\natexlab{}.
\newblock \showarticletitle{Portfolio: finding relevant functions and their usage}. In \bibinfo{booktitle}{\emph{Proceedings of the 33rd International Conference on Software Engineering, {ICSE} 2011, Waikiki, Honolulu , HI, USA, May 21-28, 2011}}. \bibinfo{publisher}{{ACM}}, \bibinfo{pages}{111--120}.
\newblock


\bibitem[Nashid et~al\mbox{.}(2024)]%
        {nashid2024contextual}
\bibfield{author}{\bibinfo{person}{Noor Nashid}, \bibinfo{person}{Taha Shabani}, \bibinfo{person}{Parsa Alian}, {and} \bibinfo{person}{Ali Mesbah}.} \bibinfo{year}{2024}\natexlab{}.
\newblock \showarticletitle{Contextual API Completion for Unseen Repositories Using LLMs}.
\newblock \bibinfo{journal}{\emph{arXiv preprint arXiv:2405.04600}} (\bibinfo{year}{2024}).
\newblock


\bibitem[Papaioannou and Doudali(2024)]%
        {papaioannou2024importance}
\bibfield{author}{\bibinfo{person}{Konstantinos Papaioannou} {and} \bibinfo{person}{Thaleia~Dimitra Doudali}.} \bibinfo{year}{2024}\natexlab{}.
\newblock \showarticletitle{The Importance of Workload Choice in Evaluating LLM Inference Systems}. In \bibinfo{booktitle}{\emph{Proceedings of the 4th Workshop on Machine Learning and Systems}}. \bibinfo{pages}{39--46}.
\newblock


\bibitem[Peng et~al\mbox{.}(2022)]%
        {peng2022revisiting}
\bibfield{author}{\bibinfo{person}{Yun Peng}, \bibinfo{person}{Shuqing Li}, \bibinfo{person}{Wenwei Gu}, \bibinfo{person}{Yichen Li}, \bibinfo{person}{Wenxuan Wang}, \bibinfo{person}{Cuiyun Gao}, {and} \bibinfo{person}{Michael~R Lyu}.} \bibinfo{year}{2022}\natexlab{}.
\newblock \showarticletitle{Revisiting, benchmarking and exploring api recommendation: How far are we?}
\newblock \bibinfo{journal}{\emph{IEEE Transactions on Software Engineering}} \bibinfo{volume}{49}, \bibinfo{number}{4} (\bibinfo{year}{2022}), \bibinfo{pages}{1876--1897}.
\newblock


\bibitem[Peng et~al\mbox{.}(2023)]%
        {DBLP:conf/kbse/PengWWGL23}
\bibfield{author}{\bibinfo{person}{Yun Peng}, \bibinfo{person}{Chaozheng Wang}, \bibinfo{person}{Wenxuan Wang}, \bibinfo{person}{Cuiyun Gao}, {and} \bibinfo{person}{Michael~R. Lyu}.} \bibinfo{year}{2023}\natexlab{}.
\newblock \showarticletitle{Generative Type Inference for Python}. In \bibinfo{booktitle}{\emph{38th {IEEE/ACM} International Conference on Automated Software Engineering, {ASE} 2023, Luxembourg, September 11-15, 2023}}. \bibinfo{publisher}{{IEEE}}, \bibinfo{pages}{988--999}.
\newblock


\bibitem[Rahman et~al\mbox{.}(2016)]%
        {rahman2016rack}
\bibfield{author}{\bibinfo{person}{Mohammad~Masudur Rahman}, \bibinfo{person}{Chanchal~K Roy}, {and} \bibinfo{person}{David Lo}.} \bibinfo{year}{2016}\natexlab{}.
\newblock \showarticletitle{Rack: Automatic api recommendation using crowdsourced knowledge}. In \bibinfo{booktitle}{\emph{2016 IEEE 23rd International Conference on Software Analysis, Evolution, and Reengineering (SANER)}}, Vol.~\bibinfo{volume}{1}. IEEE, \bibinfo{pages}{349--359}.
\newblock


\bibitem[Rawte et~al\mbox{.}(2023)]%
        {rawte2023survey}
\bibfield{author}{\bibinfo{person}{Vipula Rawte}, \bibinfo{person}{Amit Sheth}, {and} \bibinfo{person}{Amitava Das}.} \bibinfo{year}{2023}\natexlab{}.
\newblock \showarticletitle{A survey of hallucination in large foundation models}.
\newblock \bibinfo{journal}{\emph{arXiv preprint arXiv:2309.05922}} (\bibinfo{year}{2023}).
\newblock


\bibitem[Rozi{\`{e}}re et~al\mbox{.}(2023)]%
        {roziere2023code}
\bibfield{author}{\bibinfo{person}{Baptiste Rozi{\`{e}}re}, \bibinfo{person}{Jonas Gehring}, \bibinfo{person}{Fabian Gloeckle}, \bibinfo{person}{Sten Sootla}, \bibinfo{person}{Itai Gat}, \bibinfo{person}{Xiaoqing~Ellen Tan}, \bibinfo{person}{Yossi Adi}, \bibinfo{person}{Jingyu Liu}, \bibinfo{person}{Tal Remez}, \bibinfo{person}{J{\'{e}}r{\'{e}}my Rapin}, \bibinfo{person}{Artyom Kozhevnikov}, \bibinfo{person}{Ivan Evtimov}, \bibinfo{person}{Joanna Bitton}, \bibinfo{person}{Manish Bhatt}, \bibinfo{person}{Cristian Canton{-}Ferrer}, \bibinfo{person}{Aaron Grattafiori}, \bibinfo{person}{Wenhan Xiong}, \bibinfo{person}{Alexandre D{\'{e}}fossez}, \bibinfo{person}{Jade Copet}, \bibinfo{person}{Faisal Azhar}, \bibinfo{person}{Hugo Touvron}, \bibinfo{person}{Louis Martin}, \bibinfo{person}{Nicolas Usunier}, \bibinfo{person}{Thomas Scialom}, {and} \bibinfo{person}{Gabriel Synnaeve}.} \bibinfo{year}{2023}\natexlab{}.
\newblock \showarticletitle{Code Llama: Open Foundation Models for Code}.
\newblock \bibinfo{journal}{\emph{CoRR}}  \bibinfo{volume}{abs/2308.12950} (\bibinfo{year}{2023}).
\newblock


\bibitem[Shrivastava et~al\mbox{.}(2023)]%
        {shrivastava2023repository}
\bibfield{author}{\bibinfo{person}{Disha Shrivastava}, \bibinfo{person}{Hugo Larochelle}, {and} \bibinfo{person}{Daniel Tarlow}.} \bibinfo{year}{2023}\natexlab{}.
\newblock \showarticletitle{Repository-level prompt generation for large language models of code}. In \bibinfo{booktitle}{\emph{International Conference on Machine Learning}}. PMLR, \bibinfo{pages}{31693--31715}.
\newblock


\bibitem[Svyatkovskiy et~al\mbox{.}(2020)]%
        {svyatkovskiy2020intellicode}
\bibfield{author}{\bibinfo{person}{Alexey Svyatkovskiy}, \bibinfo{person}{Shao~Kun Deng}, \bibinfo{person}{Shengyu Fu}, {and} \bibinfo{person}{Neel Sundaresan}.} \bibinfo{year}{2020}\natexlab{}.
\newblock \showarticletitle{Intellicode compose: Code generation using transformer}. In \bibinfo{booktitle}{\emph{Proceedings of the 28th ACM joint meeting on European software engineering conference and symposium on the foundations of software engineering}}. \bibinfo{pages}{1433--1443}.
\newblock


\bibitem[Tang et~al\mbox{.}(2023)]%
        {tang2023domain}
\bibfield{author}{\bibinfo{person}{Ze Tang}, \bibinfo{person}{Jidong Ge}, \bibinfo{person}{Shangqing Liu}, \bibinfo{person}{Tingwei Zhu}, \bibinfo{person}{Tongtong Xu}, \bibinfo{person}{Liguo Huang}, {and} \bibinfo{person}{Bin Luo}.} \bibinfo{year}{2023}\natexlab{}.
\newblock \showarticletitle{Domain Adaptive Code Completion via Language Models and Decoupled Domain Databases}. In \bibinfo{booktitle}{\emph{2023 38th IEEE/ACM International Conference on Automated Software Engineering (ASE)}}. IEEE, \bibinfo{pages}{421--433}.
\newblock


\bibitem[Touvron et~al\mbox{.}(2023)]%
        {touvron2023llama}
\bibfield{author}{\bibinfo{person}{Hugo Touvron}, \bibinfo{person}{Louis Martin}, \bibinfo{person}{Kevin Stone}, \bibinfo{person}{Peter Albert}, \bibinfo{person}{Amjad Almahairi}, \bibinfo{person}{Yasmine Babaei}, \bibinfo{person}{Nikolay Bashlykov}, \bibinfo{person}{Soumya Batra}, \bibinfo{person}{Prajjwal Bhargava}, \bibinfo{person}{Shruti Bhosale}, \bibinfo{person}{Dan Bikel}, \bibinfo{person}{Lukas Blecher}, \bibinfo{person}{Cristian Canton{-}Ferrer}, \bibinfo{person}{Moya Chen}, \bibinfo{person}{Guillem Cucurull}, \bibinfo{person}{David Esiobu}, \bibinfo{person}{Jude Fernandes}, \bibinfo{person}{Jeremy Fu}, \bibinfo{person}{Wenyin Fu}, \bibinfo{person}{Brian Fuller}, \bibinfo{person}{Cynthia Gao}, \bibinfo{person}{Vedanuj Goswami}, \bibinfo{person}{Naman Goyal}, \bibinfo{person}{Anthony Hartshorn}, \bibinfo{person}{Saghar Hosseini}, \bibinfo{person}{Rui Hou}, \bibinfo{person}{Hakan Inan}, \bibinfo{person}{Marcin Kardas}, \bibinfo{person}{Viktor Kerkez}, \bibinfo{person}{Madian Khabsa},
  \bibinfo{person}{Isabel Kloumann}, \bibinfo{person}{Artem Korenev}, \bibinfo{person}{Punit~Singh Koura}, \bibinfo{person}{Marie{-}Anne Lachaux}, \bibinfo{person}{Thibaut Lavril}, \bibinfo{person}{Jenya Lee}, \bibinfo{person}{Diana Liskovich}, \bibinfo{person}{Yinghai Lu}, \bibinfo{person}{Yuning Mao}, \bibinfo{person}{Xavier Martinet}, \bibinfo{person}{Todor Mihaylov}, \bibinfo{person}{Pushkar Mishra}, \bibinfo{person}{Igor Molybog}, \bibinfo{person}{Yixin Nie}, \bibinfo{person}{Andrew Poulton}, \bibinfo{person}{Jeremy Reizenstein}, \bibinfo{person}{Rashi Rungta}, \bibinfo{person}{Kalyan Saladi}, \bibinfo{person}{Alan Schelten}, \bibinfo{person}{Ruan Silva}, \bibinfo{person}{Eric~Michael Smith}, \bibinfo{person}{Ranjan Subramanian}, \bibinfo{person}{Xiaoqing~Ellen Tan}, \bibinfo{person}{Binh Tang}, \bibinfo{person}{Ross Taylor}, \bibinfo{person}{Adina Williams}, \bibinfo{person}{Jian~Xiang Kuan}, \bibinfo{person}{Puxin Xu}, \bibinfo{person}{Zheng Yan}, \bibinfo{person}{Iliyan Zarov}, \bibinfo{person}{Yuchen
  Zhang}, \bibinfo{person}{Angela Fan}, \bibinfo{person}{Melanie Kambadur}, \bibinfo{person}{Sharan Narang}, \bibinfo{person}{Aur{\'{e}}lien Rodriguez}, \bibinfo{person}{Robert Stojnic}, \bibinfo{person}{Sergey Edunov}, {and} \bibinfo{person}{Thomas Scialom}.} \bibinfo{year}{2023}\natexlab{}.
\newblock \showarticletitle{Llama 2: Open Foundation and Fine-Tuned Chat Models}.
\newblock \bibinfo{journal}{\emph{CoRR}}  \bibinfo{volume}{abs/2307.09288} (\bibinfo{year}{2023}).
\newblock


\bibitem[treesitter(2024)]%
        {treesitter}
\bibfield{author}{\bibinfo{person}{treesitter}.} \bibinfo{year}{2024}\natexlab{}.
\newblock \bibinfo{title}{treesitter}.
\newblock \bibinfo{howpublished}{\url{https://github.com/tree-sitter/tree-sitter}}.
\newblock


\bibitem[Wang et~al\mbox{.}(2022)]%
        {wang2022no}
\bibfield{author}{\bibinfo{person}{Chaozheng Wang}, \bibinfo{person}{Yuanhang Yang}, \bibinfo{person}{Cuiyun Gao}, \bibinfo{person}{Yun Peng}, \bibinfo{person}{Hongyu Zhang}, {and} \bibinfo{person}{Michael~R Lyu}.} \bibinfo{year}{2022}\natexlab{}.
\newblock \showarticletitle{No more fine-tuning? an experimental evaluation of prompt tuning in code intelligence}. In \bibinfo{booktitle}{\emph{Proceedings of the 30th ACM joint European software engineering conference and symposium on the foundations of software engineering}}. \bibinfo{pages}{382--394}.
\newblock


\bibitem[Wei et~al\mbox{.}(2022)]%
        {wei2022clear}
\bibfield{author}{\bibinfo{person}{Moshi Wei}, \bibinfo{person}{Nima~Shiri Harzevili}, \bibinfo{person}{Yuchao Huang}, \bibinfo{person}{Junjie Wang}, {and} \bibinfo{person}{Song Wang}.} \bibinfo{year}{2022}\natexlab{}.
\newblock \showarticletitle{Clear: contrastive learning for api recommendation}. In \bibinfo{booktitle}{\emph{Proceedings of the 44th International Conference on Software Engineering}}. \bibinfo{pages}{376--387}.
\newblock


\bibitem[Wei et~al\mbox{.}(2023)]%
        {DBLP:journals/corr/abs-2312-02120}
\bibfield{author}{\bibinfo{person}{Yuxiang Wei}, \bibinfo{person}{Zhe Wang}, \bibinfo{person}{Jiawei Liu}, \bibinfo{person}{Yifeng Ding}, {and} \bibinfo{person}{Lingming Zhang}.} \bibinfo{year}{2023}\natexlab{}.
\newblock \showarticletitle{Magicoder: Source Code Is All You Need}.
\newblock \bibinfo{journal}{\emph{CoRR}}  \bibinfo{volume}{abs/2312.02120} (\bibinfo{year}{2023}).
\newblock


\bibitem[Xie et~al\mbox{.}(2023)]%
        {DBLP:journals/corr/abs-2305-04764}
\bibfield{author}{\bibinfo{person}{Zhuokui Xie}, \bibinfo{person}{Yinghao Chen}, \bibinfo{person}{Chen Zhi}, \bibinfo{person}{Shuiguang Deng}, {and} \bibinfo{person}{Jianwei Yin}.} \bibinfo{year}{2023}\natexlab{}.
\newblock \showarticletitle{ChatUniTest: a ChatGPT-based automated unit test generation tool}.
\newblock \bibinfo{journal}{\emph{CoRR}}  \bibinfo{volume}{abs/2305.04764} (\bibinfo{year}{2023}).
\newblock


\bibitem[Xu et~al\mbox{.}(2023)]%
        {xu2023wizardlm}
\bibfield{author}{\bibinfo{person}{Can Xu}, \bibinfo{person}{Qingfeng Sun}, \bibinfo{person}{Kai Zheng}, \bibinfo{person}{Xiubo Geng}, \bibinfo{person}{Pu Zhao}, \bibinfo{person}{Jiazhan Feng}, \bibinfo{person}{Chongyang Tao}, \bibinfo{person}{Qingwei Lin}, {and} \bibinfo{person}{Daxin Jiang}.} \bibinfo{year}{2023}\natexlab{}.
\newblock \showarticletitle{WizardLM: Empowering large pre-trained language models to follow complex instructions}. In \bibinfo{booktitle}{\emph{The Twelfth International Conference on Learning Representations}}.
\newblock


\bibitem[Zhou et~al\mbox{.}(2021)]%
        {zhou2021boosting}
\bibfield{author}{\bibinfo{person}{Yu Zhou}, \bibinfo{person}{Xinying Yang}, \bibinfo{person}{Taolue Chen}, \bibinfo{person}{Zhiqiu Huang}, \bibinfo{person}{Xiaoxing Ma}, {and} \bibinfo{person}{Harald Gall}.} \bibinfo{year}{2021}\natexlab{}.
\newblock \showarticletitle{Boosting API recommendation with implicit feedback}.
\newblock \bibinfo{journal}{\emph{IEEE Transactions on Software Engineering}} \bibinfo{volume}{48}, \bibinfo{number}{6} (\bibinfo{year}{2021}), \bibinfo{pages}{2157--2172}.
\newblock


\end{thebibliography}

\end{document}